\documentstyle[12pt,latex8,psfig]{article}          

\newcommand{\beq}{\begin{equation}} 
\newcommand{\eeq}{\end{equation}}
\newcommand{\bea}{\begin{eqnarray*}}
\newcommand{\eea}{\end{eqnarray*}}

\newcommand{\ord}{{\cal O}}
\newcommand{\cho}{\left( \! \sub \begin{array}{c}}
\newcommand{\ose}{\end{array} \! \right)}
\newcommand{\mat}{\left( \! \sub \begin{array}{cc}}
\newcommand{\Mat}{\left( \! \sub \begin{array}{cccc}}
\newcommand{\bigmat}{\left( \! \sub \begin{array}{cccccccc}}
\newcommand{\rix}{\end{array} \! \right)}
\newcommand{\sub}{\scriptsize}
\newcommand{\id}{{\bf 1}}

\newcommand{\mket}{\rangle}

\def\be{\begin{equation}}
\def\ee{\end{equation}}
\def\bea{\begin{eqnarray}}
\def\eea{\end{eqnarray}}

\gdef\makemath#1{\ifmmode #1 \else $ #1 $\fi}
\newcommand{\ignore}[1]{}

\newwrite\noteFile%
\immediate\openout\noteFile=\jobname.notes%

\def\saveNote#1{%
\immediate\write\noteFile{Page \expandafter\thepage; }%
\immediate\write\noteFile{\expandafter#1}%
\immediate\write\noteFile{}%
}

\def\authNote#1{}

\def\QED{\hfill\fbox{\phantom{.}}}

\newtheorem{theorem}{Theorem}[section]
\newtheorem{thm}{Theorem}[section]

\newtheorem{cor}[theorem]{Corollary}
\newtheorem{lemma}[theorem]{Lemma}
\newtheorem{lem}[theorem]{Lemma}

\newtheorem{defn}[theorem]{Definition}

\newtheorem{proposition}[theorem]{Proposition}
\newtheorem{prop}[theorem]{Proposition}

{\xdef\kindOfTheorem{#1}\begin{\kindOfTheorem}}
{\end{\kindOfTheorem}}%

{\par\medskip\noindent{\bf #1}\begin{it}}%
{\end{it}\par\medskip}

\def\compclassfont#1{{\sf{#1}}}

\def\NP{\compclassfont{NP}}

\def\P{\compclassfont{P}}
\def\poly{\compclassfont{poly}}

\def\PP{\compclassfont{PP}}

\def\QNP{\compclassfont{NQP}}

\def\co-C{\compclassfont{co-C}}

\def\CequalsP{\compclassfont{C_=P}}
\def\coCequalsP{\compclassfont{co}\CequalsP}

\newcommand{\ket}[1]{{|{#1} \rangle}}
\newcommand{\bra}[1]{{\langle {#1}|}}

\newcommand{\nums}{{\bf N}}
\newcommand{\nats}{{\nums}}
\newcommand{\ints}{{\bf Z}}
\newcommand{\rats}{{\bf Q}}

\newcommand{\complexes}{{\bf C}}

\newcommand{\NC}{\compclassfont{NC}}
\newcommand{\QNC}{\compclassfont{QNC}}

\newcommand{\AC}{\compclassfont{AC}}
\newcommand{\ACC}{\compclassfont{ACC}}
\newcommand{\QAC}{\compclassfont{QAC}}
\newcommand{\QACwf}{\compclassfont{QAC}_{\rm wf}}
\newcommand{\QACC}{\compclassfont{QACC}}
\newcommand{\QACCwf}{\compclassfont{QACC}_{\rm wf}}
\newcommand{\TC}{\compclassfont{TC}}
\newcommand{\QTC}{\compclassfont{QTC}}
\newcommand{\QACCP}{\QACC_{pl}^{\log}}
\newcommand{\QACCG}{\QACC_{gates}^{\log}}
\newcommand{\BQACC}{\compclassfont{BQACC}}
\newcommand{\NQACC}{\compclassfont{NQACC}}
\newcommand{\NQTC}{\compclassfont{NQTC}}
\newcommand{\EQACC}{\compclassfont{EQACC}}
\newcommand{\BQACCG}{\BQACC_{\rats,gates}^{\log}}
\newcommand{\NQACCG}{\NQACC_{gates}^{\log}}
\newcommand{\EQACCG}{\EQACC_{gates}^{\log}}
\newcommand{\BQACCP}{\BQACC_{\rats,pl}^{\log}}
\newcommand{\NQACCP}{\NQACC_{pl}^{\log}}
\newcommand{\EQACCP}{\EQACC_{pl}^{\log}}

\newcommand{\Mod}{{\rm Mod}}
\newcommand{\MOD}{{\rm MOD}}

\newenvironment{proof}%
{\medskip

\noindent {\bf Proof.}
}
{\QED \medskip}

\newcommand{\monus}{\mathbin{\mathchoice%
{\buildrel .\lower.6ex\hbox{\vphantom{.}} \over {\smash-}}%
{\buildrel .\lower.6ex\hbox{\vphantom{.}} \over {\smash-}}%
{\buildrel .\lower.4ex\hbox{\vphantom{.}} \over {\smash-}}%
{\buildrel .\lower.3ex\hbox{\vphantom{.}} \over {\smash-}}}}

\newcommand{\GN}[1]{\,\!^{\lceil}\!#1\,\!^{\rceil}}

\newcommand{\AND}{\wedge}

\newcommand{\HALF}[1]{\lfloor\frac{1}{2}#1\rfloor}

\begingroup\makeatletter\ifx\SetFigFont\undefined
\def\x#1#2#3#4#5#6#7\relax{\def\x{#1#2#3#4#5#6}}%
\expandafter\x\fmtname xxxxxx\relax \def\y{splain}%
\ifx\x\y   
\gdef\SetFigFont#1#2#3{%
  \ifnum #1<17\tiny\else \ifnum #1<20\small\else
  \ifnum #1<24\normalsize\else \ifnum #1<29\large\else
  \ifnum #1<34\Large\else \ifnum #1<41\LARGE\else
     \huge\fi\fi\fi\fi\fi\fi

  \csname #3\endcsname}%
\else
\gdef\SetFigFont#1#2#3{\begingroup
  \count@#1\relax \ifnum 25<\count@\count@25\fi
  \def\x{\endgroup\@setsize\SetFigFont{#2pt}}%
  \expandafter\x
    \csname \romannumeral\the\count@ pt\expandafter\endcsname
    \csname @\romannumeral\the\count@ pt\endcsname
  \csname #3\endcsname}%
\fi
\fi\endgroup

\begin{document}

\title{Counting, Fanout, and the Complexity of Quantum ACC}
\author{
Frederic Green\\
Department of Mathematics and Computer Science\\
Clark University, Worcester, MA 01610\\
fgreen@black.clarku.edu\\
\and
Steven Homer~\thanks{Supported in part by the NSF
under grant NSF-CCR-9988310}\\
Computer Science Department\\ 
Boston University, Boston, MA 02215\\
homer@cs.bu.edu\\
\and
Cristopher Moore~\thanks{Supported in part by the NSF
under grant NSF-PHY-0071139}\\
Computer Science Department\\
University of New Mexico, Albuquerque NM 87131\\
and the Santa Fe Institute\\
moore@cs.unm.edu\\
\and
Christopher Pollett\\
Department of Mathematics\\
University of California, Los Angeles, CA\\
cpollett@math.ucla.edu\\
}
\maketitle

\thispagestyle{empty}

\begin{abstract}
We propose definitions of $\QAC^0$, the quantum analog of the
classical class $\AC^0$ of constant-depth circuits with AND and OR
gates of arbitrary fan-in, and $\QACC[q]$, the analog of the class
$\ACC[q]$ where $\Mod_q$ gates are also allowed.  We prove that parity
or fanout allows us to construct quantum $\MOD_q$ gates in constant
depth for any $q$, so $\QACC[2] = \QACC$.  More generally, we show
that for any $q,p > 1$, $\MOD_q$ is equivalent to $\MOD_p$ (up to
constant depth).  This implies that $\QAC^0$ with unbounded fanout
gates, denoted $\QACwf^0$, is the same as $\QACC[q]$ and $\QACC$ for
all $q$.  Since $\ACC[p] \ne \ACC[q]$ whenever $p$ and $q$ are
distinct primes, $\QACC[q]$ is strictly more powerful than its
classical counterpart, as is $\QAC^0$ when fanout is allowed.  This
adds to the growing list of quantum complexity classes which are
provably more powerful than their classical counterparts.

We also develop techniques for proving upper bounds for $\QACC^0$ in
terms of related language classes. We define classes of languages
$\EQACC$, $\NQACC$ and $\BQACC_{\rats}$. We define a notion of
$\log$-planar $\QACC$ operators and show the appropriately restricted
versions of $\EQACC$ and $\NQACC$ are contained in $\P/\poly$. We also
define a notion of $\log$-gate restricted $\QACC$ operators and show
the appropriately restricted versions of $\EQACC$ and $\NQACC$ are
contained in $\TC^0$.
\end{abstract}

\section{Introduction}

Advances in quantum computation in the last decade have been
among the most notable in theoretical computer
science. This is due to the surprising improvements in
the efficiency of solving several fundamental  combinatorial problems
using quantum mechanical methods in place  of their classical counterparts.
These advances led to considerable efforts in finding
new efficient
quantum algorithms for
classical problems and  in developing a complexity
theory of quantum computation.

While most of the original results in quantum computation were
developed using quantum Turing machines,
they can also be formulated in terms of quantum circuits, which
yield a more natural model of quantum
computation. For example, Shor
\cite{shor97} has shown that quantum circuits can  factor integers more
efficiently
than
any known classical algorithm for factoring.  And quantum circuits have
been shown
(see Yao
\cite{yao93}) to provide a universal model for quantum computation.

The theory of circuit complexity has long been an important branch of
theoretical computer science.  Shallow circuits correspond to parallel
algorithms that can be performed in small amounts of time on a
massively parallel computer with constant communication delays, and so
circuit complexity can be thought of as a study of how to solve
problems in parallel.  In addition, some low-lying circuit classes
have beautiful algebraic characterizations, e.g.\
\cite{barrington,beaudry,circuits}.

In \cite{codes,early}, Moore and Nilsson suggested a definition of
$\QNC$, the quantum analog of the class $\NC$ of problems solvable by
circuits with polylogarithmic depth and polynomial size \cite{papa}.
Here, we will study quantum versions of some additional circuit
classes.  Recall the following definitions:
\begin{enumerate}
\item $\NC^{k}$ consists of problems solvable by families of circuits of
AND, OR, and NOT gates with depth $\ord(\log^k n)$ and size polynomial
in $n$, where $n$ is the size of the input, and where the AND and OR
gates have just two inputs each.

\item $\AC^{k}$ is like $\NC^k$, but where we allow AND and OR gates
with unbounded fan-in, i.e.\ arbitrary numbers of inputs, in each layer
of the circuit.

\item $\ACC^{k}[q]$ is like $\AC^k$, but where we also allow $\Mod_q$ gates
with unbounded fan-in, where $\Mod_q(x_1,\ldots,x_n)$ outputs 1 iff
the sum of the inputs is not a multiple of $q$.

\item $\ACC^{k} = \cup_q \,\ACC^{k}[q]$.

\item $\NC = \cup_k \,\NC^{k} = \cup_k \,\AC^{k} = \cup_k \,\ACC^{k}$.
\end{enumerate}
Then we have
\[ \AC^{0} \subset \ACC^{0}[2] \subset \ACC^{0} \subseteq \NC^{(1)}
\subseteq \cdots
     \subseteq \NC \]
In fact, these first two inclusions are known to be proper
\cite{ajtai,furst,razborov,smo87}.  Neither {\sc Majority} nor
{\sc Parity} are in $\AC^{0}$, while the latter is trivially in
$\ACC^{0}[2]$.  In addition, $\ACC^{0}[p]$ and $\ACC^{0}[q]$ are known to
be
incomparable whenever $p$ and $q$ are distinct primes.  Thus these
classes give us some of the few strict inclusions known in
computational complexity theory.  However, for all anyone knows,
$\ACC^{0}[6]$ could contain $\PP$, $\NP$, and the entire polynomial
hierarchy!

Quantum analogs of $\AC^0$ and $\ACC$ are defined and studied here.
One central class that we examine is
a quantum analog of $\AC^{0}$ that we denote
$\QACwf^{0}$.
$\QACwf^{0}$ is the class of families of operators which
can be built out of products of constantly many layers 
consisting of polynomial-sized tensor
products of one-qubit gates (analogous to NOT's), Toffoli gates (analogous
to AND's and OR's) and
fan-out gates. The subscript ``$wf$" in the notation denotes ``with
fan-out." The idea of fan-out in the quantum setting is subtle, as is
made clear in Section~\ref{CatStateSection} of this paper. The sub-class
of $\QACwf^0$ that does not include fan-out gates is denoted simply
$\QAC^0$.
An analog of \ACC$[q]$ (i.e., \ACC\/ circuit families only allowing
Mod$_q$ gates) is 
$\QACC[q]$, defined similarly to $\QACwf^{0}$, but
replacing the fan-out gates with
quantum $\Mod_q$ gates (which we denote as $\MOD_q$). The class $\QACC$
is $\cup_q \QACC[q]$.

In this paper, we prove a number of results about $\QAC$ and $\QACC$,
and address some definitional difficulties.  We show that an ability
to form a ``cat state'' with $n$ qubits, or fan out a qubit into $n$
copies in constant depth, is equivalent to being able to construct an
$n$-ary parity gate in constant depth.  We discuss how best to compare
these
circuit classes to classical ones.

We  prove the surprising result  that, for any integer $q > 1$,
$\QACwf^{0} = \QACC[q] = \QACC$. This is in sharp contrast to
the classical result of Smolensky~\cite{smo87} that says $\ACC^{0}[q] 
\not = \ACC^{0}[p]$ for any pair of distinct primes $q, p$, which implies
that for any prime $p$, $\AC^{0} \subset \ACC^{0}[p] \subset 
\ACC$. 
This result shows that parity gates are as powerful as any other
mod gates in $\QACC$, 
and more generally, that any $\MOD_q$ gate is as good as any other,
up to polynomial size and constant depth.
 Thus we conclude that $\QACwf^0$, or, for any $q$, $\QACC[q]$, 
is strictly more powerful than
$\ACC[q]$ and $\AC^{0}$.
 

  We also develop methods for proving upper bounds for $\QACC$.
The definition of $\QACC$ immediately leads to a problem in this regard:
$\QACC$ is a class of operators that only have a natural interpretation
quantum mechanically. In order to clarify the
relationship with classical
computation we assign properties to $\QACC$ circuits based
on measurements we can perform on them.
In particular, we define several natural languages classes
related to $\QACC$.
These language classes arise from considering a quantum circuit family in 
the class and 
specifying a condition on the expectation
of observing a particular state after applying a
circuit from the family to an input state. The condition might
simply be that the expectation is non-zero, or that it is bounded away from
zero by some constant, or that it is exactly equal to some constant. 
We call the language classes obtained by these conditions on the
expectation
$\NQACC$, $\BQACC$ and $\EQACC$, respectively.
For example, the class $\NQACC$ corresponds to the
case where $x$ is in the language if the expectation of the observed
state after applying the $\QACC$ operator is non-zero.
This  is analogous to
the definition of
the class $\QNP$ as defined in Adleman et al.~\cite{ADH97}
and discussed in Fenner et
al.~\cite{fghr98}.
In this way we obtain  natural classes of languages which correspond to
those
defined classically by families of small depth circuits.
In these terms, for example, we can more succinctly and precisely
express the
statement ``$\QACC[q]$ is strictly more powerful than $\ACC[q]$'' by
writing $\ACC[q] \subset \EQACC[q]$.

We 
desire upper bounds showing that 
these language classes are contained in classically defined circuit
classes, thus delimiting the power of these quantum computations.
In particular, we believe that the languages arising in this way 
from our definitions
are contained within $\TC^{0}$, those problems computed by
constant-depth threshold circuits. We have been unable to verify this,
and in fact the only classical upper bound for these 
language classes that we know of is the very powerful counting class
$\coCequalsP$
(see~\cite{fghr98}).  We do give some evidence for this proposed $\TC^{0}$
upper bound here
and further provide some techniques which may prove useful in solving this
problem.

Our methods result in upper bounds for restricted
$\QACC$ circuits. Roughly speaking, we show that $\QACC$ is no more
powerful than $\P/\poly$
provided that a layer of ``wire-crossings" in the $\QACC$ operator can be
written as log
many compositions of Kronecker  products of  controlled-not gates. We call
this class
$\QACCP$,  where the ``pl'' is for this planarity condition. We show if one
further
restricts attention to the case where the number of multi-line gates 
(gates whose input is more than 1 qubit) is
log-bounded then the circuits are no
more powerful than $\TC^{0}$. We call this class $\QACCG$.
These results hold for arbitrary complex amplitudes in the $\QACC$
circuits.

In terms of our language classes, we show that $\NQACCG$\, is in 
$\TC^{0}$ and $\NQACCP$\, is in $\P/\poly$. Although the proof uses some of
the 
techniques developed by 
Fenner, Green, Homer and Pruim ~\cite{fghr98} and by
Yamakami and
Yao~\cite{yy98} to show that  $\QNP_{C} = \coCequalsP$,
the small depth circuit case presents technical challenges
not present in their setting. In particular, given a $\QACC$ operator built
out
of layers $M_1, \ldots , M_t$ and an input state $\ket{x, 0^{p(n)}}$, we 
must show that a $\TC^{0}$ circuit can keep track of the
amplitudes of each possible resulting state as each layer is applied. 
After all layers have been applied, the $\TC^{0}$ circuit then needs to be
able to check that the amplitude of one possible state is
non-zero. Unfortunately, there
could be exponentially many states with non-zero amplitudes after applying 
a 
layer. 
To handle this problem we introduce the idea of a
``tensor-graph," a new way to 
represent a collection of states. We can extract from these graphs 
(via  $\TC^{0}$ or $\P/\poly$ computations) 
whether the amplitude of any particular vector 
is 
non-zero.

The exponential growth in the number of states is one of the
primary obstacles to proving that all of $\NQACC$ is in $\TC^{0}$
(or even $\P/\poly$), and thus the tensor graph formalism represents
a significant step
towards such an upper bound. The reason the bounds apply only
in the restricted cases is that although tensor graphs
can represent any $\QACC$ operator, in the case of
operators with layers that might do arbitrary permutations, 
the top-down approach we use to compute a desired amplitude from the graph 
no longer seems to work. We feel that it is likely 
that the amplitude of any vector in a tensor graph can be written
as a polynomial product of a polynomial sum in some extension algebra
of the ones we work with in this paper, in which case it is quite
likely it can be evaluated in $\TC^{0}$.

Another important obstacle to obtaining a $\TC^{0}$ upper bound is
that one needs to
be able to add and multiply a 
polynomial number of complex amplitudes
that may appear in a $\QACC$ computation.
We solve this problem.
It reduces to adding and multiplying polynomially many
elements of a
certain transcendental extension of the rational numbers. We
show that in fact
$\TC^{0}$ is closed under iterated addition and multiplication
of such numbers (Lemma~\ref{sumprod} below). This result is of independent 
interest, 
and our application of tensor-graphs and these closure properties of 
$\TC^{0}$ may prove useful in further 
investigations of small-depth quantum circuits.

We now discuss the organization of the rest of this paper. 
Section~2 contains definitions for the quantum operator classes we will
be considering as well as other background definitions. Section~3
shows the constant-depth quantum circuit equivalence of fan-out and parity
gates. Section~4 establishes for arbitrary $p$ and $q$ the constant-depth
quantum equivalence of $\Mod_p$ and $\Mod_q$. Section~5 contains our upper
bound results. Finally, the last section has a conclusion and some open
problems. 

Preliminary versions of these results appeared in~\cite{moore99} and
\cite{ghp00}.


\section{Preliminaries}

In this section we define the gates used as building blocks for our
quantum circuits. Classes of operators built out of these gates are
then defined. We define language classes that can be determined by
these operators and give a couple of definitions from algebra. Lastly,
some closure properties of $\TC^0$ are described.

\begin{defn}\label{gate-definitions}
We define various quantum gates as follows:

\begin{itemize}

\item By a {\em one-qubit gate} we mean an operator from the group $U(2)$.

\item 
Let $U = \left( \begin{array}{cc} u_{00} & u_{01} \\ u_{10} & u_{11}
\end{array} \right) \in U(2)$. $\AND_m(U)$ is defined as: $\AND_0(U)=U$ 
and for $m>0$, $\AND_m(U)$ is
\[ \AND_m(U)(\ket{\vec{x},y}) = \left\{
\begin{array}{ll}
u_{y0}\ket{\vec{x},0} + u_{y1}\ket{\vec{x},1} & \mbox{if }
\AND^m_{k=1}x_k=1\\
\ket{\vec{x},y} & \mbox{otherwise}
\end{array} \right. \]

\item Let $X = \sigma_x = \left( \begin{array}{cc} 0 & 1 \\ 1 & 0
\end{array} \right)$.  A {\em Toffoli gate} is a $\AND_m(X)$ gate for
some $m \geq 0$.  A {\em controlled-not} gate is a $\AND_1(X)$ gate.

\item The {\em Hadamard gate} is the one-qubit gate $H =
\frac{1}{\sqrt{2}} \left( \begin{array}{cc} 1 & 1 \\ 1 & -1
\end{array} \right)$.

\item An {\em (m-)spaced controlled-not gate} is an operator that
maps $\ket{y_1,\ldots, y_m,x}$ to $\ket{x\oplus y_1, y_2\ldots, y_m,x}$
or $\ket{x,y_1,\ldots, y_m}$ to $\ket{x, y_1\ldots, y_{m-1},y_m\oplus x}$  

\item An {\em  (m-ary) fan out gate} $F$ is an operator that maps
$\ket{y_1,\ldots, y_m,x}$ to $\ket{x\oplus y_1,\ldots, x\oplus y_m,x}$.
\item The classical Boolean $\Mod_q$-function on $n$ bits is defined
so that $\Mod_q(x_1,\ldots,x_n) = 1$ ${\rm iff}$ $\sum_{i=1}^n x_i
\not \equiv 0 \bmod{q}.$ We also define $\Mod_{q,r}(x_1,...,x_n)$ to
output 1 iff $\sum_{i=1}^n x_i \equiv r \bmod{q}$. A quantum $\MOD_q$
gate is an operator that maps $\ket{y_1,\ldots, y_m,x}$ to
$\ket{y_1,\ldots, y_m,x \oplus \Mod_q(y_1,\ldots, y_m)}$.  A quantum
$\MOD_{q,r}$ gate maps $\ket{y_1,\ldots, y_m,x}$ to $\ket{y_1,\ldots,
y_m,x \oplus \Mod_{q,r}(y_1,\ldots, y_m)}$.  We write $\neg \MOD_q$
for $\MOD_{q,0}$.  A {\em parity} gate is a $\MOD_2$ gate.

\end{itemize}
\end{defn}

\begin{figure}
\centerline{\psfig{file=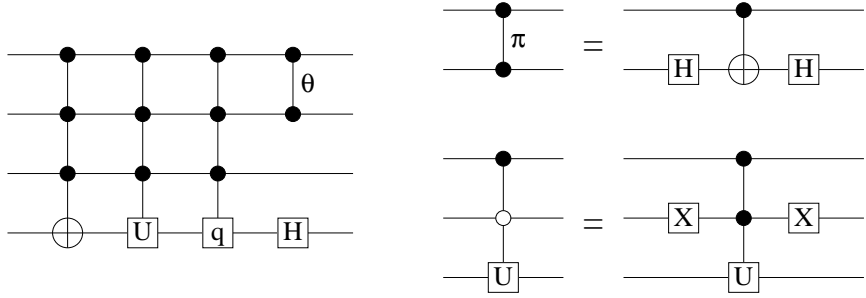,width=4.5in}}
\caption{Our notation for $n$-ary Toffoli, controlled-$U$, and
$\MOD_q$ gates, fanout gates, symmetric phase shift gates, and the
Hadamard gate.  On the top right, we show a useful identity between
the controlled-not, the controlled $\pi$-shift, and the Hadamard gate.
On the bottom right we show a controlled-$U$ gate with one of its
inputs negated by conjugation with $X$.}
\label{notation}
\end{figure}

Note that, since negation is built into the output (via the exclusive
OR), it is easy to simulate negations using $\MOD_{q,r}$ gates (unlike
the classical case).  For example, by setting $b=1$, we can compute
$\neg\Mod_{q,r}$. More generally, using one work bit, it is possible
to simulate ``$\neg\MOD_{q,r}$," defined so that,
\[ \ket{x_1,...,x_n,b} \mapsto
\ket{x_1,...,x_n,b\oplus(\neg\Mod_{q,r}(x_1,...,x_n))}
\]
using just $\MOD_{q,r}$ and a controlled-not gate. Thus $\MOD_{q,r}$ and
$\neg\MOD_{q,r}$ are equivalent up to constant depth. 
Finally, observe that $\MOD^{-1}_{q,r} = \MOD_{q,r}$.

We will use the notation in Figure~\ref{notation} for our various gates.

As discussed in further detail in section~\ref{CatStateSection} below,
the no-cloning theorem of quantum mechanics makes it difficult to
directly fan out qubits in constant depth (although constant fan-out
in constant depth is no problem, since we can make multiple
copies of the inputs). Thus it is necessary to define the operator $F$
as in the above definition.  Also, in the literature it is frequently
the case that one says a given operator $M$ on $\ket{y_1, \ldots,
y_m}$ can be written as a tensor product of certain gates $M_j$.  What
is meant is that there is an permutation operator $\Pi$ ( a map from
$\ket{y_1, \ldots, y_m}$ to $\ket{y_{\pi(1)},\ldots, y_{\pi(m)}}$ for
some permutation $\pi$) such that
\[
M\ket{y_1, \ldots y_m} = \Pi\otimes^n_j M_j \Pi^{-1}\ket{y_1, \ldots
y_m}
\]
where the $M_j$'s are our base gates, i.e.\ those gates for which no
inherent ordering on the $y_i$ is assumed {\it a priori}, and
$\otimes$ is the Kronecker product, which flattens a tensor product
into a matrix with blocks indexed in a particular way.  Since it is
important to keep track of such details in our upper bounds proofs, we
will always use Kronecker products of the form $\otimes^n_j M_j$
without unspoken permutations.  Nevertheless, being able to do
permutation operators (not conjugation by a permutation) intuitively
allows our circuits to simulate classical wire crossings. To handle
permutations, we allow our circuits to have controlled-not layers. A
{\em controlled-not layer} is a gate which performs, in one step,
controlled-not's between an arbitrary collection of disjoint pairs of
lines in its domain. That is, it performs $\Pi \otimes^n_j \AND_1(X)
\Pi^{-1}$ for some permutation operator $\Pi$.  It is easy to
see~\cite{codes} that any permutation can be written as a product of
a constant number of controlled-not layers. We say a controlled-not
layer is {\em log-depth} if it can be written as the composition of
log many matrices each of which is the Kronecker product of identities
and spaced controlled-not gates.

$M^{\otimes n}$ is the $n$-fold Kronecker product of $M$ with itself.

\begin{defn}\label{qacc-definition}
\item $\QAC^k$ is the class of families $\{F_n\}$, where $F_n$ is in
$U(2^{n+p(n)})$, $p$ a polynomial, and each $F_n$ is writable as a product
of $\ord(\log^k n)$  layers,  where a {\em layer} is a Kronecker product of
one-qubit gates and Toffoli gates or is a controlled-not layer. Also for
all $n$  the number of distinct types of one qubit gates used must be
fixed. 
\item $\QACC^k[q]$ is the same as $\QAC^k$ except we also
allow $\MOD_q$ gates.  $\QACC^k = \cup_q \QACC^k[q]$.

\item $\QAC^k_{wf}$ is the same as $\QAC^k$ but we also allow
fan-out gates.

\item $\QACC$ is defined as $\QACC^0$ and $\QACC[q]$ is defined as
$\QACC^0[q]$.  $\QACCP$ is $\QACC$ restricted to log-depth controlled
not layers.  $\QACCG$ is $\QACC$ restricted so that the total number
of multi-line gates in all layers is log-bounded.

\item If ${\mathcal C}$ is one of the above classes and $K \subseteq
\complexes$, then ${\mathcal C}_K$ are the families in ${\mathcal C}$
with coefficients restricted to $K$.

\item Let $\{F_n\}$ and $\{G_n\}$, $G_n, F_n\in U(2^{n})$ be families of
operators. We say $\{F_n\}$ is {\rm $\QAC^0$ reducible}
to $\{G_n\}$ if there is a family
$\{R_n\}$, $R_n\in U(2^{n+p(n)})$ of $\QAC^0$ operators augmented with
operators from $\{G_n\}$ such that for all $n$, ${\bf x},{\bf y}
\in \{0,1\}^n$, there is a setting of $z_1,...,z_{p(n)} \in \{0,1\}$
for which $\bra{{\bf y}}F_n\ket{{\bf x}} = \bra{{\bf y}, {\bf z}}R_n
\ket{{\bf x},{\bf z}}$. Operator families are {\rm $\QAC^0$ equivalent}
if they are $\QAC^0$ reducible to each other. If ${\cal C}_1$
and ${\cal C}_2$ are families of $\QAC^0$ equivalent operators, we write
${\cal C}_1 = {\cal C}_2$.
\end{defn}

We refer to the $z_i$'s above as ``work bits" (also called ``ancillae"
in~\cite{codes}).  Note that in proving $\QAC^0$ equivalence, the
work bits must be returned to their original values in a computation
so that they are disentangled from the rest of the circuit, and can be
re-used by subsequent layers.

It follows for any $\{F_n\} \in \QAC^0$ that $F_n$ is writable as a
product of finite number of layers.  In an earlier paper,
Moore~\cite{moore99} places no restriction on the number of distinct
types of one-qubit gates used in a given family of operators.  Here we
restrict these so that the number of distinct amplitudes which appear
in matrices in a layer is fixed with respect to $n$.  This restriction
arises implicitly in the quantum Turing machine case of the upper
bounds proofs in Fenner, et al.~\cite{fghr98} and Yamakami and
Yao~\cite{yy98}. Also, it seems fairly natural since in the classical
case one builds circuits using a fixed number of distinct gate
types. Our classes here are, thus, more ``uniform'' than those defined
earlier ~\cite{moore99}.  We now define language classes based on our
classes of operator families.

\begin{defn}
Let ${\mathcal C}$ be a class of families of $U(2^{n+p(n)})$ operators
where $p$ is a polynomial and $n=|x|$.
\begin{enumerate}
\item E$\cdot\mathcal{C}$ is the class of languages $L$ such that for
some $\{F_n\} \in \mathcal{C}$ and
$\{\bra{\vec{z}_n}\}=\{\bra{z_{n,1}, \ldots, z_{n,n+p(n)}}\}$ a family
of states, $m:=|\bra{\vec{z}_n}F_n\ket{x, 0^{p(n)}}|^2$ is $1$ or $0$
and $x\in L$ iff $m=1$.
\item N$\cdot\mathcal{C}$ is the class of languages $L$ such that for some 
$\{F_n\} \in \mathcal{C}$ and $\{\bra{\vec{z}_n}\}$ a family of states,
$x\in L$ iff $|\bra{\vec{z}_n}F_n\ket{x, 0^{p(n)}}|^2>0$.
\item B$\cdot\mathcal{C}$ is the class of languages $L$ where for 
$\{F_n\} \in \mathcal{C}$ and $\{\bra{\vec{z}}\}$,
$x\in L$ if $|\bra{\vec{z}_n}F_n\ket{x, 0^{p(n)}}|^2>3/4$ and $x\not\in 
L$ 
if $|\bra{\vec{z}_n}F_n\ket{x, 0^{p(n)}}|^2<1/4$ .  
\end{enumerate}
\end{defn}

It follows E$\cdot\mathcal{C}\subseteq$ N$\cdot\mathcal{C}$ and
E$\cdot\mathcal{C}\subseteq$ B$\cdot\mathcal{C}$. We frequently will
omit the `$\cdot$' when writing a class, so E$\cdot$\QACC\, is written
as \EQACC. Let $\ket{\Psi} := F_n\ket{x, 0^{p(n)}}$.  Notice that
$|\bra{\vec{z}_n}F_n\ket{x, 0^{p(n)}}|^2 =
\bra{\Psi}P_{\ket{\vec{z}_n}}\ket{\Psi}$, where $P_{\ket{\vec{z}_n}}$
is the projection matrix onto $\ket{\vec{z}_n}$. We could allow in our
definitions measurements of up to polynomially many such projection
observables and not affect our results below.  However, this would
shift the burden of the computation in some sense away from the
$\QACC$ operator and instead onto preparation of the observable.

Next are some variations on familiar definitions from algebra.

\begin{defn}
Let $k>0$. A subset $\{\beta_i\}_{1\leq i\leq k}$ of $\complexes$ is
{\em linearly independent} if $\sum^k_{i=1} a_i\beta_i \neq 0$ for any
$(a_1,\ldots, a_k) \in \rats^k - \{\vec{0}^k\}$.  A set
$\{\beta_i\}_{1\leq i\leq k}$ is {\em algebraically independent} if
the only $p\in\rats[x_1,\ldots,x_k]$ with
$p(\beta_1,\ldots,\beta_k)=0$ is the zero polynomial. \end{defn}

We now briefly mention some closure properties of $\TC^0$ computable
functions that are useful in proving $\NQACCG \subseteq \TC^0$.  For
proofs of the statements in the next lemma
see~\cite{siu91,siu94,clote93}.

\begin{lemma}
(1) $\TC^0$ functions are closed under composition. (2) The following
are $\TC^0$ computable: $x+y$, $x\monus y:= x - y$ if $x-y > 0$ and
$0$ otherwise, $|x| := \lceil \log_2(x + 1 )\rceil$, $x\cdot y$,
$\lfloor x/y \rfloor$, $2^{\min(i,p(|x|)}$, and $cond(x,y,z) := y$ if
$x>0$ and $z$ otherwise. (3) If $f(i,x)$ is $\TC^0$ computable then
$\sum^{p(|x|)}_{k=0} f(k,x)$, $\prod^{p(|x|)}_{k=0} f(k,x)$, $\forall
i \leq p(|x|)(f(i,x)=0)$, $\exists i \leq p(|x|)(f(i,x)=0)$, and
$\mu_{i \leq p(|x|)}(f(i,x)=0):=$ the least $i$ such that $f(i,x)=0$
and $i \leq p(|x|)$ or $p(x)+1$ otherwise, are $\TC^0$ computable.
\end{lemma}
We drop the $\min$ from the $2^{\min(i,p(|x|))}$ when it is obvious a
suitably large $p(|x|)$ can be found. We define $max(x,y):=
cond(1\monus (y\monus x)),x,y)$ and define
\begin{eqnarray*}
max_{i\leq p(|x|)}(f(i))  := 
 \mu_{i \leq  p(|x|)} (\forall j \leq p(|x|)(f(j)\monus f(i) = 0)
\end{eqnarray*}
Using the above functions we describe a way to do sequence coding in
$\TC^0$.  Let $\beta_{|t|}(x,w) := \lfloor(w \monus \lfloor
w/2^{(x+1)|t|}\rfloor\cdot2^{(x+1)|t|})/2^{x|t|}\rfloor.$ The function
$\beta_{|t|}$ is useful for block coding. Roughly, $\beta_{|t|}$ first
gets rid of the bits after the $(x+1)|t|$th bit then chops off the low
order $x|t|$ bits. Let $B= 2^{|\max(x,y)|}$, so that $B$ is longer
than either $x$ or $y$.  Hence, we code pairs as $\langle x,y \rangle
:= (B+y)\cdot 2B + B+x$, and projections as $(w)_1 :=
\beta_{\HALF{|w|}\monus 1}(0, \beta_{\HALF{|w|}}(0,w))$ and $(w)_2 :=
\beta_{\HALF{|w|}\monus 1}(0,\beta_{\HALF{|w|}}(1,w))$. We can encode
a poly-length, $\TC^0$ computable sequence of numbers $\langle f(1),
\ldots, f(k) \rangle$ as the pair $\langle \sum^k_i(f(i)2^{i\cdot m}),
m\rangle$ where $m:=|f(\max_i(f(i)))|+1$. We then define the function
which projects out the $i$th member of a sequence as $\beta(i,w) :=
\beta_{(w)_2}(i,w)$.

We can code integers using the positive natural numbers by letting the
negative integers be the odd natural numbers and the positive integers
be the even natural numbers. $\TC^0$ can use the $\TC^0$ circuits for
natural numbers to compute both the polynomial sum and polynomial
product of a sequence of $\TC^0$ definable integers. It can also
compute the rounded quotient of two such integers. For instance, to do
a polynomial sum of integers, compute the natural number which is the
sum of the positive numbers in the sum using $cond$ and our natural
number iterated addition circuit. Then compute the natural number
which is the sum of the negative numbers in the sum. Use the
subtraction circuit to subtract the smaller from the larger number and
multiply by two.  One is then added if the number should be
negative. For products, we compute the product of the natural numbers
which results by dividing each integer code by two and rounding
down. We multiply the result by two. We then sum the number of terms
in our product which were negative integers. If this number is odd we
add one to the product we just calculated. Finally, division can be
computed using the Taylor expansion of $1/x$.

\section{Fanout, Cat States, and Parity}\label{CatStateSection}

To make a shallow parallel circuit, it is often important to {\em fan
out} one of the inputs into multiple copies.  One of the differences
between classical circuits and quantum ones as we have defined them
here is that in classical circuits, we usually assume that we get
arbitrary fanout for free, simply by splitting a wire into as many
copies as we like.  This is difficult in quantum circuits, since
making an unentangled copy requires non-unitary, and in fact
non-linear, processes:
\[ (\alpha |0\mket + \beta |1\mket) \otimes (\alpha |0\mket + \beta
|1\mket) 
 = \alpha^2 |00\mket + \alpha \beta (|01\mket + |10\mket) + \beta^2
|11\mket\ 
\]
has coefficients quadratic in $\alpha$ and $\beta$, so it cannot be
derived from $\alpha |0\mket + \beta |1\mket$ using any linear
operator, let alone a unitary one.  This is one form of the so-called
``no cloning'' theorem.

However, the controlled-not gate can be used to copy a qubit onto a 
work bit
in the pure state $|0\mket$ by making a non-destructive measurement:
\[ (\alpha |0\mket + \beta |1\mket ) \otimes |0\mket \,\to\,
   \alpha |00\mket + \beta |11\mket \]

Note that the final state is not a tensor product of two independent 
qubits,
since the two qubits are completely entangled.  This means that whatever
we do to one copy, we do to the other.  Except when the states are purely
Boolean, we have to treat this kind of ``fanout'' more gingerly than
we would in the classical case.

\begin{figure}
\centerline{\psfig{figure=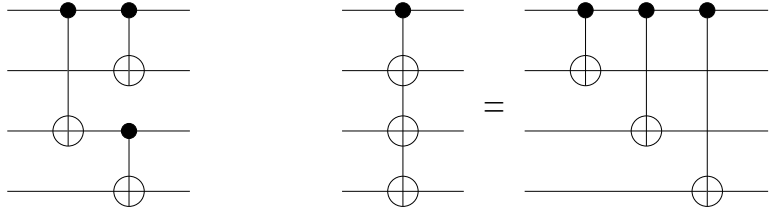,width=4in}}
\caption{Two ways to make a cat state on $n$ qubits.  The circuit on the
left uses only two-qubit gates and has depth $\log n$.  On the right, we
define a ``fanout gate'' that simultaneously performs $n$ controlled-nots
from one input qubit.}
\label{fanout}

\end{figure}

By making $n$ copies of a qubit in this sense, we can make a ``cat
state'' $\alpha |000 \cdots 0\mket + \beta |111 \cdots 1\mket$.  Such
states are useful in making quantum computation fault-tolerant (e.g.\
\cite{ds,shor2}).  We can do this in $\log n$ depth with
controlled-not gates, as shown on the left-hand side of
Figure~\ref{fanout}.  When preceded by a Hadamard gate on the top
qubit, this circuit will map an initial state $|0000\mket$ onto a cat
state $\frac{1}{\sqrt{2}}(|0000\mket + |1111\mket)$.  However, we will
also consider circuits which can do this in a single layer, with a
``fanout gate'' that simultaneously copies a qubit onto $n$ target
qubits.  This is simply the product of $n$ controlled-not gates, as
shown on the right-hand side of Figure~\ref{fanout}.

We now show that in quantum circuits, we can do fanout in constant depth
if and only if we can construct a parity gate in constant depth.

\begin{proposition}  In any class of quantum circuits that includes
Hadamard and controlled-not gates, the following are equivalent:
\begin{enumerate}
\item It is possible to map $\alpha|0\mket + \beta|1\mket$ and $n-1$
work bits in the state $|0\mket$ onto an $n$-qubit cat state
$\alpha |000 \cdots 0\mket + \beta |111 \cdots 1\mket$ in constant depth.

\item The $n$-ary fanout gate on the right-hand side of
Figure~\ref{fanout} can be implemented in constant depth with at most
$n-1$ additional work bits.

\item An $n$-ary parity or $\MOD_2$ gate as defined above can 
be implemented
in constant depth with at most $n-1$ additional work bits.
\end{enumerate}
\label{fanoutprop}
\end{proposition}

\begin{figure}
\centerline{\psfig{file=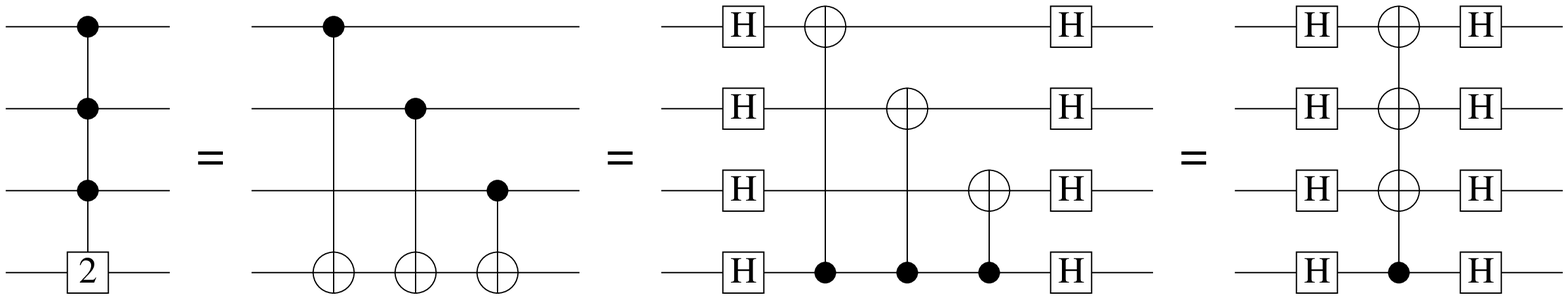,width=5in}}
\caption{The parity and fanout gates are conjugates of each other by a
layer of Hadamard gates.}
\label{conj}
\end{figure}

\begin{figure}
\centerline{\psfig{file=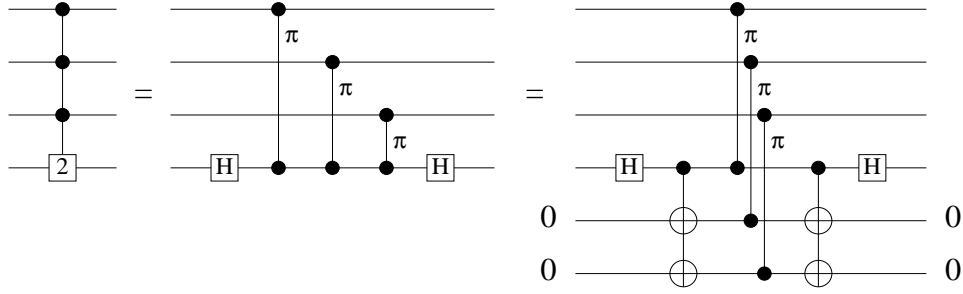,width=5in}}
\caption{The parity gate can also be written as a product of controlled
$\pi$-shifts, with the target qubit conjugated by $H$.  Since these are
diagonal, we can parallelize them using any gate that can make a cat state\
.}
\label{mod2}
\end{figure}

\begin{proof}  First, note that (1) is {\em a priori} weaker than (2),
since (1) only requires that an operator map $|100 \cdots 0\mket$ to
$|111 \cdots 1\mket$ and $|000 \cdots 0\mket$ to itself.  In fact, the
two circuits shown in Figure~\ref{fanout} both do this, even though
they differ on other initial states.

To prove $(2 \Leftrightarrow 3)$, we simply need to notice that the
parity gate is a fanout gate going the other way conjugated by a layer
of Hadamard gates, since parity is simply a product of controlled-nots
with the same target qubit, and conjugating with $H$ reverses the direction
of a controlled-not.  This is shown in Figure~\ref{conj}.  Clearly the
number of work bits used to perform either gate will be the same. (We
prove this equivalence in greater detail and generality in
Proposition~\ref{conjugates} below.)

To prove $(1 \Rightarrow 3)$, we use a slightly more elaborate circuit
shown in Figure~\ref{mod2}.  Here we use the identity shown in
Figure~\ref{notation} to convert the parity gate into a product of
controlled $\pi$-shifts.  Since these are diagonal, they can be
parallelized as in \cite{codes} by copying the target qubit onto $n-1$
work bits, and applying each one to a different copy.  While we have
drawn the circuit with two fanout gates, any gate that satisfies the
conditions in (1), and its inverse after the $\pi$-shifts, will do.

Finally, $(2 \Rightarrow 1)$ is obvious.
\end{proof}

This brings up an interesting issue.  It is not clear that a $\QAC^0$
operator as we have defined $\QAC^0$ here can ``simulate'' any $AC^0$
circuit, since we are not allowing arbitrary fanout in each layer.  An
alternate definition, which we might call {\em $\QAC$ with fanout} or
$\QACwf^0$, would allow us to perform controlled-$U$ gates or Toffoli
gates in the same layer whenever they have different target qubits,
even if their input qubits overlap.  This seems reasonable, since
these gates commute.  Since we can fan out to $n$ copies in $\log n$
layers as in Figure~\ref{fanout}, we have $\QAC^k \subseteq
\QACwf^k \subseteq \QAC^{(k+1)}$.  We can define $\QACCwf$ in the
same way, and Proposition~\ref{fanoutprop} implies that $\QACwf^k
= \QACCwf^k[2] = \QACC^k[2]$.

It is partly a matter of taste whether $\QAC^0$ or $\QACwf^0$ is a
better analog of $\AC^0$.  However, fanout does seem possible in
several proposed technologies for quantum computing.  In an ion trap
computer \cite{cirac}, vibrational modes can couple with all the atoms
simultaneously, so we could apply a controlled-not from one atom to
the ``bus qubit'' and then from the bus to the other $n$ atoms.  In
bulk-spin NMR \cite{gershenfeld}, we can activate the couplings from
one atom to $n$ others, and perform $n$ controlled $\pi$-shifts
simultaneously, which is equivalent to fanout with the target qubits
conjugated with the Hadamard gate.  Thus allowing fanout may in fact
be the most reasonable model of constant-depth quantum circuits.

\section{Constant Depth Equivalence of $\MOD_p$ and $\MOD_q$ Gates}
 
As stated in the Introduction, in the classical case $\Mod_p$ 
and $\Mod_q$ gates are not
easy to build from each other whenever $p$ and $q$ are relatively prime.
In fact, to do it in constant depth requires a circuit of exponential size
\cite{smo87}.  In this section, we will show this is not true in the
quantum case.  Specifically, we show that any $\MOD_q$ gate can be
built in constant depth from any $\MOD_p$ gate, for any two numbers
$p$ and $q$. We start by showing that any $\MOD_q$ gate can be built
from parity gates in constant depth. 
\begin{proposition}\label{q-reduces-to-2}
  In any circuit class containing $n$-ary parity gates
and one-qubit gates, we can construct an $n$-ary $\MOD_q$ gate, with
$\ord(n \,\log q)$ work bits, in depth depending only on $q$.
\end{proposition}

\begin{proof}  Let $k = \lceil \log_2 q \rceil$, 
and let $M$ be a Boolean
matrix on $k$ qubits where the zero state has\
 period $q$.  For
instance, if we write $|x\mket$ as shorthand for $|x_{k-1} \cdots x_1
x_0\mket$ where $x_i$ is the $2^i$ digit of $x$'s binary expansion and
$0 \le x < 2^k$, we can define $M$ so that it\
 permutes the $|x\mket$ as
follows:
\[ M |x\mket = \left\{ \begin{array}{ll}
       |(x+1) \bmod q\mket & \quad \mbox{if } \
x < q \\ |x\mket & \quad
       \mbox{if } x \ge q \end{array} \right.\
 \]

Then if we start with $k$ work bits in the state $|0\mket$ and apply a
controlled-$M$ gate to them from each input, the state will differ from
 $|0\mket$ on at least one qubit if and only if the number of true
inputs is not a multiple of $q$.  (Note that this controlled-$M$ gate
applies to $k$ target qubits at once in an entangled way.)  We can
then apply an $n$-ary OR of
these $k$ qubits to the target qubit,
i.e. a Toffoli gate with its inputs conjugated with $X$ and its target
qubit negated before or after the gate.
  We end by applying the
inverse series of controlled-$M^\dagger$ gates to return the $k$
work bits to $|0\mket$.

Now we use Proposition~4 of \cite{codes} to parallelize this set of
controlled-$M$ gates.  We can convert them to diagonal gates by
conjugating the $k$ qubits with a unitary operator $T$, where
$T^\dagger D T = M$ and $D$ is diagonal.  If we have a parity gate, we
can fan out the $k$ work bits to $n$ copies each using
Proposition~\ref{fanoutprop}.  We can then simultaneously apply the
$n$ controlled-$D$ gates from each input to the corresponding copy,
and then uncopy them back.

This is shown in Figure~\ref{mod3}.  For $q=3$, for instance, 
$M = \Mat 0 & 1 & 0 &   \\
          0 & 0 & 1 &   \\
          1 & 0 & 0 &   \\
            &   &   & 1 \rix$,
$T = \frac{1}{\sqrt{3}}
     \Mat & & & \sqrt{3} \\
          1 & 1 & 1 &   \\
          e^{4\pi i/3} & e^{2\pi i/3} & 1 & \\
          e^{2\pi i/3} & e^{4\pi i/3} & 1 & \rix$,
and $D = \Mat 1 & & & \\
          & 1 & & \\
          & & e^{2\pi i/3} & \\
          & & & e^{4\pi i/3} \rix$.

\begin{figure}
\centerline{\psfig{file=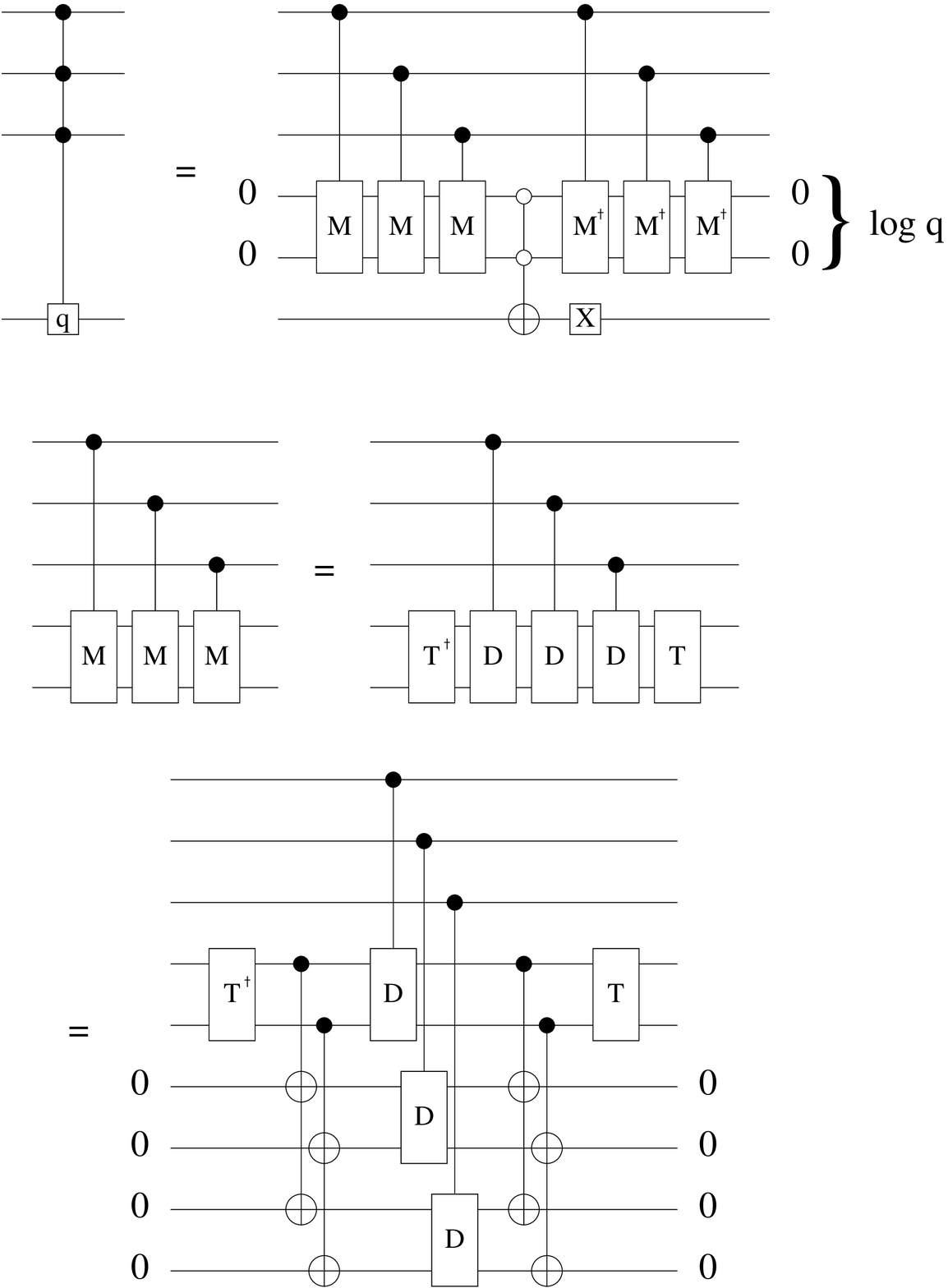,width=4.5in}}
\caption{Building a $\Mod_q$ gate.  We choose a matrix $M$ on $k =
\lceil \log_2 q \rceil$ qubits such that $M^q = \id$, apply
controlled-$M$ gates from the $n$ inputs to $k$ work bits, apply an OR
from these $k$ qubits to the target qubit, and reverse the process to
return the work bits to $|0\mket$.  To parallelize this, we can
diagonalize $M$ by writing it as $T^\dagger D T$, fan the $k$ qubits
out into $n$ copies each using Proposition~\ref{fanoutprop}, and apply
controlled-$D$ gates simultaneously from each input to a set of
copies.  The total depth depends on $q$ but not on $n$.}
\label{mod3}
\end{figure}

The operators $T$, $T^\dagger$, and the controlled-$D$ gate can be
carried out in some finite depth by controlled-nots and one-qubit
gates by the results of \cite{barenco95}.  The total depth of our
$\MOD_q$ gate is a function of these and so of $q$, but not of $n$.
Finally, the number of work bits used is $(n-1)k = \ord(n \,\log q)$
as promised.
\end{proof}

To look more closely at the depth as a function of $q$, we note that
using the methods of Reck et al.\ \cite{reck} and Barenco et
al.~\cite{barenco95}, any operator on $k$ qubits can be performed with
$\ord(k^3 \,4^k)$ two-qubit gates.  Since $k = \lceil \log_2 q
\rceil$, this means that the depths of $T$, $T^\dagger$ and the
controlled-$D$ gates are at most $\ord(q^2 \log^3 q)$.

Since we can construct $\MOD_q$ gates in constant depth, we have
$\QACC^k[q] \subset \QACC^k[2]$ for all $q$, so $\QACC^k =
\QACC^k[2]$.  By Proposition~\ref{fanoutprop}, these are both also
equal to $\QACwf^k$.  In particular, we have
\[ \QACwf^0 = \QACC[2] = \QACC \]
while classically both equalities are strict inclusions.  Note that
allowing fanout immediately gives $\QACCwf[q] = \QACC[2] = \QACC$ for
any $q$, but we will show below that including fanout explicitly is
not necessary.

We now show the converse $\QACC[2] \subseteq \QACC[q]$ for any $q$,
i.e.\ parity can be built from $\MOD_q$ for any $q$.  This shows the
constant depth equivalence of $\MOD_q$ gates for all $q$.

Let $q \in \nums$, $q \ge 2$ be fixed for the remainder of this
section. Consider quantum states labeled by digits in $D =
\{0,...,q-1\}$. By analogy with ``qubit," we refer to a state of the
form,
\[ \sum\limits_{k=0}^{q-1} c_k\ket{k} \]
with $\sum_k |c_k|^2 = 1$ as a ``qudigit."

\begin{figure}[ht]
\begin{center}
\begin{picture}(2400,1550)
\put(750,500){\line(-1,0){300}}
\put(950,500){\line(1,0){300}}
\put(450,800){\line(1,0){800}}
\put(450,1400){\line(1,0){800}}
\put(850,1400){\line(0,-1){800}}
\put(850.25,800.25){\circle*{100}}
\put(850.25,1400.25){\circle*{100}}
\put(150,1350){\makebox(250,100)[r]{$x_1$}}
\put(750,400){\framebox(200,200)[]{$q$}}
\put(100,750){\makebox(300,100)[r]{$x_n$}}
\put(1300,1350){\makebox(300,100)[l]{$x_1$}}
\put(1300,750){\makebox(300,100)[l]{$x_n$}}
\put(100,450){\makebox(300,100)[r]{$b$}}
\put(1300,450){\makebox(1000,100)[l]{$(b + x_1 + ... + x_n)~{\rm mod}~q$}}
\put(500,1250){\makebox(100,100)[]{.}}
\put(500,1100){\makebox(100,100)[]{.}}
\put(500,950){\makebox(100,100)[]{.}}
\put(1100,1250){\makebox(100,100)[]{.}}
\put(1100,1100){\makebox(100,100)[]{.}}
\put(1100,950){\makebox(100,100)[]{.}}
\end{picture}
\end{center}
  \caption{An $M_q$ gate.}
\label{modAddOperator}
\end{figure}
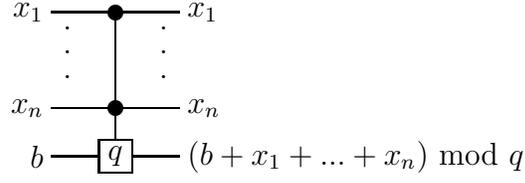

We define three important operations on qudigits. The $n$-ary {\it modular
addition} operator $M_q$ acts as follows:
\[ 
M_q \ket{x_1,...,x_n, b} = \ket{x_1,...x_n,
(b + x_1 + ... + x_n)\bmod q}
\]
We use the same graphical notation for $M_q$ as we do for a $\MOD_q$ gate,
but
interpreting the lines as qudigits, as illustrated in
Figure~\ref{modAddOperator}.

Since $M_q$ merely permutes the states, it is clear that it is unitary.
Similarly, the $n$-ary unitary {\it base $q$ fanout} operator $F_q$ acts
as,
\[ F_q \ket{x_1,...x_n,b} = \ket{(x_1 + b)\bmod q,...(x_n + b)\bmod q,b} \]
We write $F$ for $F_2$, since it is the ``standard" fan-out gate
introduced in Definition~\ref{gate-definitions}.  Note that $M^{-1}_q
= M_q^{q-1}$ and $F^{-1}_q = F_q^{q-1}$.

Finally, the Quantum Fourier Transform $H_q$ (which generalizes the
Hadamard
transform $H$ on qubits) acts on a single qudigit as,
\[ H_q \ket{a} = {1\over \sqrt{q}}\sum\limits_{b=0}^{q-1} \zeta^{ab}
\ket{b}
\]
where $\zeta = e^{2\pi i \over q}$ is a primitive complex $q^{th}$ root of 
unity. It is easy to see that $H_q$ is unitary, via the fact that
$\sum_{\ell = 0}^{q-1} \zeta^{a\ell} = 0$ iff $ a \not \equiv 0\bmod q$.

 The first observation is that, analogous to parity and fanout for Boolean
inputs, the operators $M_q$ and $F_q$ are ``conjugates" in the following
sense.
This is a generalization of the equivalence of assertions (2) and (3) of
Proposition~\ref{fanoutprop}.

\begin{prop}\label{conjugates}
  $M_q = (H_q^{\otimes (n+1)})^{-1} F^{-1}_q H_q^{\otimes (n+1)}.$
\end{prop}
\begin{proof}
  We apply the operators $H_q^{\otimes (n+1)}$, $F^{-1}_q$, and
$(H_q^{\otimes (n+1)})^{-1}$ in that order to the state $\ket{x_1,...,x_n,
b}$,
and check that the result has the same effect as $M_q$.

The operator $H_q^{\otimes (n+1)}$ simply applies $H_q$ to
each of the $n+1$ qudigits of $\ket{x_1,...,x_n, b}$, which yields,
\begin{eqnarray*}
         {1 \over q^{(n+1)\over 2}}\sum\limits_{{\bf y} \in D^n}
          \sum\limits_{a = 0}^{q-1} \zeta^{{\bf x}\cdot {\bf y} + ab}
               \ket{y_1,...,y_n, a},
\end{eqnarray*}
where ${\bf y}$ is a compact notation for $y_1,...,y_n$, and
${\bf x}\cdot{\bf y}$ denotes $\sum_{i=1}^n x_iy_i$.
Then applying $F^{-1}_q$ to the above state yields,
\begin{eqnarray*}
     {1 \over q^{(n+1)\over 2}}\sum\limits_{{\bf y} \in D^n}
          \sum\limits_{a = 0}^{q-1}\zeta^{{\bf x}\cdot {\bf y} + ab}
               |(y_1-a)\bmod q,...,(y_n - a)\bmod q, a\rangle.
\end{eqnarray*}
By a change of variable, the above can be re-written as,
\begin{eqnarray*}
     {1 \over q^{(n+1)\over 2}}\sum\limits_{{\bf y} \in D^n}
          \sum\limits_{a = 0}^{q-1} \zeta^{\sum_{i=1}^n x_i(y_i+a) + ab}
               \ket{y_1,...,y_n, a}
\end{eqnarray*}
Finally, applying $(H_q^{\otimes (n+1)})^{-1}$ to the above undoes the
Fourier transform and
puts the coefficient of $a$ in the exponent into the last slot of the
state.

The result is,
\begin{eqnarray*}
  (H_q^{\otimes (n+1)})^{-1} F^{-1}_q H_q^{\otimes (n+1)}
  \ket{x_1,...,x_n,b} =
             \ket{x_1,...,x_n, (b + x_1 + ... + x_n) \bmod q},
\end{eqnarray*}
which is exactly what $M_q$ would yield.
\end{proof}

  We now describe how the operators $M_q$, $F_q$ and $H_q$ can be modified
to
operate on registers consisting of qu{\it bits} rather than qu{\it digits}.
Firstly, we encode each digit using $\lceil \log q \rceil$ bits. Thus, for
example, when $q = 3$, the basis states $\ket{0}, \ket {1}$ and $\ket{2}$
are
represented by the two-qubit registers $\ket{00}, \ket{01}$ and $\ket{10}$,
respectively. Note that there remains one state (in the example,
$\ket{11}$)
which does not correspond to any of the qudigits. In general, there will be
$2^{\lceil \log q\rceil} - q$ such ``non-qudigit" states. $M_q$, $F_q$ and
$H_q$ can now be defined to act on qubit registers, as follows. Consider a
state $\ket{x}$ where $x$ is a number represented as $m$ bits (i.e., an
$m$-qubit register). If $m < \lceil \log q\rceil$, then $H_q$ leaves
$\ket{x}$
unaffected. If $0 \le x \le q -1$ (where here we are identifying $x$ with
the
number it represents), then $H_q$ acts exactly as one expects, namely,
$ H_q\ket{x} = (1 / \sqrt{q})\sum_{y=0}^{q-1} \zeta^{xy}\ket{y}.$
If $x \ge q$, again $H_q$ leaves $\ket{x}$ unchanged. Since the resulting
transformation is a direct sum of unit matrices and matrices of the form
of $H_q$ as it was originally set down, the result is a unitary
transformation.
$M_q$ and $F_q$ can be defined to operate similarly on $m$-qubit registers
for
any $m$: Break up the $m$ bits into blocks of $\lceil \log q\rceil$ bits.
If $m$ is not divisible by $\lceil \log q \rceil$, then $M_q$ and $F_q$ do
not
affect the ``remainder" block that contains fewer than $\lceil \log q
\rceil$
bits. Likewise, in a quantum register $\ket{x_1,...,x_n}$ where each of the
$x_i$'s (with the possible exception of $x_n$)
are $\lceil \log q \rceil$-bit numbers, $M_q$
and
$F_q$ operate on the blocks of bits $x_1,...,x_n$ exactly as expected,
except that 
there is no affect on the ``non-qudigit" blocks (in which $x_i \ge q$), or
on
the (possibly) one remainder block for which $|x_n| < \lceil \log q
\rceil$.
Since $M_q$ and $F_q$ operate exactly as they did originally on blocks
representing qudigits, and like unity for non-qudigit or remainder blocks,
it is clear
that they remain unitary.

Henceforth, $M_q$, $F_q$, and $H_q$ should be understood to act on qubit
registers as described above. Nevertheless, it will usually be convenient
to
think of them as acting on qu{\it digit} registers consisting of
$\lceil\log
q\rceil$ qubits in each.

\begin{lemma}\label{F_q=M_q}
 $F_q$ and $M_q$ are $\QAC^0$-equivalent.
\end{lemma}
\begin{proof}
By Barenco et al.~\cite{barenco95}, any fixed dimension unitary matrix
can be computed in fixed depth using one-qubit gates and controlled
nots. Hence $H_q$ can be computed in $\QAC^0$, as can $H_q^{\otimes
(n+1)}$. The result now follows immediately from
Proposition~\ref{conjugates}.
\end{proof}

\begin{lemma}
$\MOD_q$ and $M_q$ are $\QAC^0$-equivalent.
\end{lemma}
\begin{proof}
First note that $\neg \MOD_q$ and $\MOD_{q,r}$ are equivalent, since
a $\MOD_{q,r}$ gate can be simulated by a $\neg \MOD_q$ gate with
$q-r$ extra inputs set to the constant 1. Since $\neg \MOD_q$ and
$\MOD_q$ gates are equivalent, we can freely use $\MOD_{q,r}$ gates in
place of $\MOD_q$ gates and vice versa.

It is easy to see that, given an $M_q$ gate, we can simulate a
$\MOD_q$ gate.  Applying $M_q$ to $n+1$ digits (represented as bits,
but each digit only taking on the values 0 or 1) transforms,
\[ \ket{x_1,...,x_n,0} \mapsto
   \ket{x_1,...,x_n,(\sum_i x_i)\bmod q}. 
\] 
Now send the bits of the last block ($\sum_i x_i\bmod q$) to an
$n$-ary OR gate with control bit $b$ (see the proof of
Proposition~\ref{q-reduces-to-2}). The resulting output is exactly $b
\oplus \Mod_q(x_1,...,x_n)$. The bits in the last block can be erased
by reversing the $M_q$ gate.  This leaves only $x_1,...,x_n$, $\ord(n)$
work bits, and the output $b \oplus \Mod_q(x_1,...,x_n)$.

The converse (simulating $M_q$ given $\MOD_q$) requires some more
work. The first step is to show that $\MOD_{q,0}$ can also determine
if a sum of {\it digits} is divisible by $q$. Let $x_1,...,x_n \in D$
be a set of digits represented as $\lceil \log q \rceil$ bits
each. For each $i$, let $x_i^{(k)}$ ($0 \le k \le \lceil \log q \rceil
- 1$) denote the bits of $x_i$.  Since the numerical value of $x_i$ is
$\sum_{k=0}^{\lceil \log q \rceil - 1} x_i^{(k)}2^k$, it follows that
\begin{eqnarray*}
  \sum\limits_{i=1}^n x_i = \sum \limits_{k=0}^{\lceil \log q \rceil - 1}
     \sum\limits_{i=1}^n 
           x_i^{(k)}2^k.
\end{eqnarray*}

The idea is to express this last sum in terms of a set of Boolean
inputs that are fed into a $\MOD_{q,0}$ gate. To account for the
factors $2^k$, each $x_i^{(k)}$ is fanned out $2^k$ times before
plugging it into the $\MOD_{q,0}$ gate. Since $k < \lceil \log q
\rceil$, this requires only constant depth and $\ord(n)$ work bits (which
of course are set back to 0 in the end by reversing the fanout).  Thus,
just using $\MOD_{q,0}$ and constant fanout, we can determine if
$\sum_{i=1}^n x_i \equiv 0 \bmod q$.  More generally, we can determine
if $\sum_{i=1}^n x_i \equiv r \bmod q$ using just a $\MOD_{q,r}$ gate
and constant fanout.  Let $\widehat{\MOD}_{q,r}(x_1,...,x_n)$ denote
the resulting circuit, that determines if a sum of digits is congruent
to $r$ mod $q$.  The construction of
$\widehat{\MOD}_{q,r}(x_1,...,x_n)$ is illustrated in
Figure~\ref{ModWideHat} for the case of $q=3$.

\begin{figure}
\begin{center}
\begin{picture}(3550,2800)
\put(3050.25,2650.25){\circle*{100}}
\put(3050.25,2350.25){\circle*{100}}
\put(3050.25,1500.25){\circle*{100}}
\put(3050.25,1200.25){\circle*{100}}
\put(3000,550){\makebox(100,100)[]{\small{$\widehat{q},r$}}}
\put(2650,2650){\line(1,0){800}}
\put(2650,2350){\line(1,0){800}}
\put(2650,1500){\line(1,0){800}}
\put(2650,1200){\line(1,0){800}}
\put(2650,600){\line(1,0){250}}
\put(3200,600){\line(1,0){250}}
\put(1900,1650){\makebox(350,250)[]{$\equiv$}}
\put(700.25,2650.25){\circle*{100}}
\put(700.25,2350.25){\circle*{100}}
\put(700.25,2050.25){\circle*{100}}
\put(700.25,1500.25){\circle*{100}}
\put(700.25,1200.25){\circle*{100}}
\put(700.25,900.25){\circle*{100}}
\put(600,500){\framebox(200,200)}
\put(650,550){\makebox(100,100)[]{$q$}}
\put(300,2650){\line(1,0){800}}
\put(300,2350){\line(1,0){800}}
\put(300,2050){\line(1,0){800}}
\put(500,2350){\line(0,-1){350}}
\put(900,2350){\line(0,-1){350}}
\put(500.25,2050.25){\circle{100}}
\put(900.25,2050.25){\circle{100}}
\put(300,1500){\line(1,0){800}}
\put(300,1200){\line(1,0){800}}
\put(300,900){\line(1,0){800}}
\put(700,2650){\line(0,-1){1950}}
\put(500,1200){\line(0,-1){350}}
\put(900,1200){\line(0,-1){350}}
\put(500.25,900.25){\circle{100}}
\put(900.25,900.25){\circle{100}}
\put(450,1850){\makebox(100,100)[]{.}}
\put(450,1700){\makebox(100,100)[]{.}}
\put(450,1550){\makebox(100,100)[]{.}}
\put(850,1850){\makebox(100,100)[]{.}}
\put(850,1700){\makebox(100,100)[]{.}}
\put(850,1550){\makebox(100,100)[]{.}}
\put(100,2600){\makebox(100,100)[]{$x_1^{(0)}$}}
\put(100,2300){\makebox(150,100)[]{$x_1^{(1)}$}}
\put(100,1400){\makebox(150,150)[]{$x_n^{(0)}$}}
\put(100,1950){\makebox(150,150)[]{0}}
\put(100,1100){\makebox(150,150)[]{$x_n^{(1)}$}}
\put(150,850){\makebox(100,100)[]{0}}
\put(300,600){\line(1,0){300}}
\put(800,600){\line(1,0){300}}
\put(100,550){\makebox(150,100)[]{$b$}}
\put(1200,2600){\makebox(100,100)[]{$x_1^{(0)}$}}
\put(1200,2300){\makebox(150,100)[]{$x_1^{(1)}$}}
\put(1200,1950){\makebox(150,150)[]{0}}
\put(1200,1400){\makebox(150,150)[]{$x_n^{(0)}$}}
\put(1200,1100){\makebox(150,150)[]{$x_n^{(1)}$}}
\put(1200,850){\makebox(100,100)[]{0}}
\put(1200,500){\makebox(930,160)[]{$b\oplus {\rm mod}(x)$}}
\put(500.25,2350.25){\circle*{100}}
\put(900.25,2350.25){\circle*{100}}
\put(500.25,1200.25){\circle*{100}}
\put(900.25,1200.25){\circle*{100}}
\put(2900,450){\framebox(300,300)}
\put(2900,600){\line(-1,0){250}}
\put(3200,600){\line(1,0){250}}
\put(3050,2650){\line(0,-1){1900}}
\end{picture}
\end{center}
\caption{ A $\widehat{\rm MOD}_{3,r}$ circuit for $r=0$.
In the figure, ${\rm mod}(x)$ denotes $\Mod_{3,r}(x_1,...,x_n)$.
The notation on the right will be used as a shorthand for
this circuit.}
\label{ModWideHat}
\end{figure}
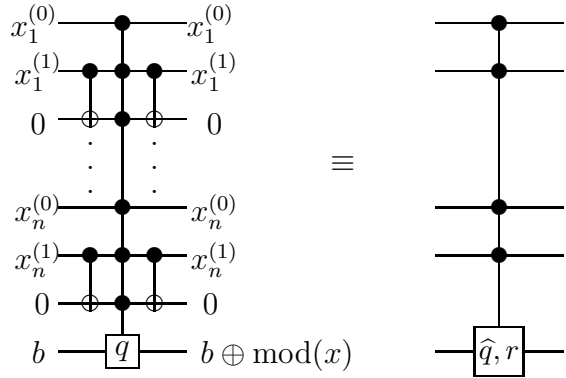

  We can get the bits in the value of the sum $\sum_{i=1}^n x_i\bmod
q$ using $\widehat{\MOD}_{q,r}$ circuits. This is done, essentially,
by implementing the relation $x \bmod q = \sum_{r=0}^{q-1}r\cdot
\Mod_{q,r}(x)$.  For each $r$, $0 \le r \le q-1$, we compute
$\Mod_{q,r}(x_1,...,x_n)$ (where now the $x_i$'s are digits). This can
be done by applying the $\widehat{\MOD}_{q,r}$ circuits in series (for
each $r$) to the same inputs, introducing a 0 work bit for each
application, as illustrated in Figure~\ref{modHatSeries}.

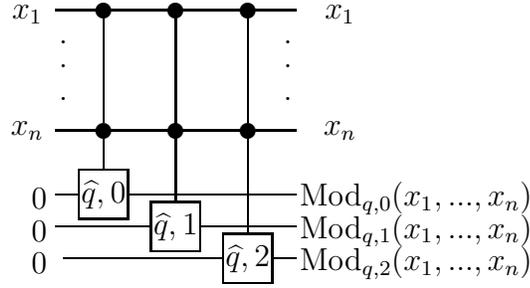
\begin{figure}[ht]
\begin{center}
\begin{picture}(2700,1950)
\put(650.25,1800.25){\circle*{100}}
\put(650.25,1050.25){\circle*{100}}
\put(1100.25,1800.25){\circle*{100}}
\put(1100.25,1050.25){\circle*{100}}
\put(1100,1750){\line(0,-1){700}}
\put(1550.25,1800.25){\circle*{100}}
\put(1550.25,1050.25){\circle*{100}}
\put(1550,1750){\line(0,-1){700}}
\put(600,600){\makebox(100,100)[]{$\widehat{q},0$}}
\put(1050,400){\makebox(100,100)[]{$\widehat{q},1$}}
\put(1500,200){\makebox(100,100)[]{$\widehat{q},2$}}
\put(350,1800){\line(1,0){1500}}
\put(350,1050){\line(1,0){1500}}
\put(2300,550){\makebox(600,150)[]{${\rm Mod}_{q,0}(x_1,...,x_n)$}}
\put(2300,350){\makebox(600,150)[]{${\rm Mod}_{q,1}(x_1,...,x_n)$}}
\put(2300,150){\makebox(600,150)[]{${\rm Mod}_{q,2}(x_1,...,x_n)$}}
\put(200,550){\makebox(100,150)[]{0}}
\put(200,350){\makebox(100,150)[]{0}}
\put(200,150){\makebox(100,150)[]{0}}
\put(100,1700){\makebox(150,150)[]{$x_1$}}
\put(2050,1700){\makebox(150,150)[]{$x_1$}}
\put(100,950){\makebox(150,200)[]{$x_n$}}
\put(2050,950){\makebox(150,200)[]{$x_n$}}
\put(350,1550){\makebox(100,100)[]{.}}
\put(350,1400){\makebox(100,100)[]{.}}
\put(350,1200){\makebox(100,100)[]{.}}
\put(1750,1550){\makebox(100,100)[]{.}}
\put(1750,1400){\makebox(100,100)[]{.}}
\put(1750,1200){\makebox(100,100)[]{.}}
\put(650,1750){\line(0,-1){650}}
\put(500,500){\framebox(300,300)}
\put(950,300){\framebox(300,300)}
\put(1400,100){\framebox(300,300)}
\put(650,1000){\line(0,-1){200}}
\put(1100,1000){\line(0,-1){400}}
\put(1550,1000){\line(0,-1){600}}
\put(500,650){\line(-1,0){150}}
\put(950,450){\line(-1,0){600}}
\put(1400,250){\line(-1,0){1000}}
\put(800,650){\line(1,0){1050}}
\put(1700,250){\line(1,0){150}}
\put(1250,450){\line(1,0){600}}
\end{picture}
\end{center}
\caption{Applying $\widehat{{\rm MOD}}_{q,r}$ circuits in series.}
\label{modHatSeries}
\end{figure}

Let $r_k$ denote the $k^{th}$ bit of $r$. 
For each $r$ and for each $k$, we take the AND of the output of the
$\widehat{\MOD}_{q,r}$ with
$r_k$ (again by applying the AND's in series, which
is still constant depth, but introduces $q$ extra work inputs). Let
$a_{k,r}$ denote the output of one of these AND's. For each
$k$,
we OR together all the $a_{k,r}$'s, that is, compute
$\vee_{r=0}^{q-1} a_{k,r}$, again introducing a constant number of
work bits. Since only
one of the $r$'s will give a non-zero output from $\widehat{\MOD}_{q,r}$,
this
collection of OR gates outputs exactly the bits in the value of
$\sum_{i=1}^n x_i \bmod q$. Call the resulting circuit $C$,
and the sum it outputs $S$.

 Finally, to simulate $M_q$, we need to include the input digit $b \in D$.
To
do this, we apply a unitary transformation $T$ to $\ket{S, b}$ that
transforms
it to $\ket{S, (b + S)\bmod q}$. 
By Barenco, et al.~\cite{barenco95}
(as in the proof of Lemma \ref{F_q=M_q}), 
$T$ can be computed in fixed depth using
one-qubit gates and controlled NOT gates.
Now using $S$ and all the other
work inputs, we reverse the computation of the circuit $C$, thus
clearing the work inputs. This is illustrated
in figure~\ref{finalM_qCircuit}.

\begin{figure}[ht]
\begin{center}
\begin{picture}(2800,1850)
\put(1200,150){\framebox(400,400)}
\put(400,450){\line(1,0){200}}
\put(900,450){\line(1,0){300}}
\put(1600,450){\line(1,0){300}}
\put(2200,450){\line(1,0){200}}
\put(400,950){\line(1,0){200}}
\put(400,1150){\line(1,0){200}}
\put(400,1650){\line(1,0){200}}
\put(2200,950){\line(1,0){200}}
\put(2200,1150){\line(1,0){200}}
\put(2200,1650){\line(1,0){200}}
\put(600,350){\framebox(300,1400)}
\put(1900,350){\framebox(300,1400)}
\put(900,1650){\framebox(1000,0)}
\put(900,1150){\framebox(1000,0)}
\put(900,950){\framebox(1000,0)}
\put(100,1550){\makebox(200,200)[]{$x_1$}}
\put(100,1050){\makebox(200,200)[]{$x_n$}}
\put(2500,1550){\makebox(200,200)[]{$x_1$}}
\put(2500,1050){\makebox(200,200)[]{$x_n$}}
\put(2300,1450){\makebox(100,100)[]{.}}
\put(2300,1350){\makebox(100,100)[]{.}}
\put(2300,1250){\makebox(100,100)[]{.}}
\put(400,1450){\makebox(100,100)[]{.}}
\put(400,1350){\makebox(100,100)[]{.}}
\put(400,1250){\makebox(100,100)[]{.}}
\put(400,750){\makebox(100,100)[]{.}}
\put(400,650){\makebox(100,100)[]{.}}
\put(400,550){\makebox(100,100)[]{.}}
\put(2300,750){\makebox(100,100)[]{.}}
\put(2300,650){\makebox(100,100)[]{.}}
\put(2300,550){\makebox(100,100)[]{.}}
\put(100,850){\makebox(200,200)[]{0}}
\put(100,350){\makebox(200,200)[]{0}}
\put(2500,850){\makebox(200,200)[]{0}}
\put(2500,350){\makebox(200,200)[]{0}}
\put(950,450){\makebox(200,200)[]{$S$}}
\put(1650,450){\makebox(200,200)[]{$S$}}
\put(1300,250){\makebox(200,200)[]{$T$}}
\put(600,950){\makebox(300,200)[]{$C$}}
\put(1900,950){\makebox(300,200)[]{$C^{-1}$}}
\put(1200,250){\line(-1,0){200}}
\put(800,100){\makebox(200,200)[]{$b$}}
\put(1600,250){\line(1,0){200}}
\put(2050,100){\makebox(800,200)[]{$(b+S)~{\rm mod}~q$}}
\end{picture}
\end{center}
\caption{Combining circuits to compute $M_q$.}
\label{finalM_qCircuit}
\end{figure}
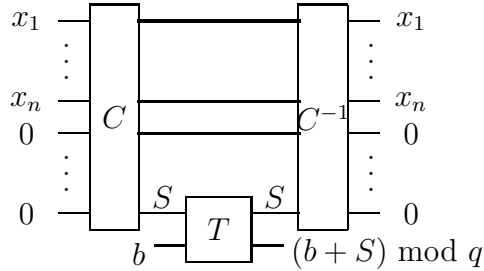

The result is an output consisting of $x_1,...,x_n$, $\ord(n)$ work
bits, and $(b + \sum_{i=1}^n x_i)\bmod q$, which is the output of an
$M_q$ gate.
\end{proof}

It is clear that we can fan out digits, and therefore bits, using an
$F_q$ gate (setting $x_i = 0$ for $1 \le i \le n$ fans out $n$ copies
of $b$). It is slightly less obvious (but still straightforward) that,
given an $F_q$ gate, we can fully simulate an $F$ gate.

\begin{lemma}\label{F-F_q-equiv}
For any $q > 2$, $F$ and $F_q$ are $\QAC^0$-equivalent.
\end{lemma}
\begin{proof} 
By the preceding lemmas, $F_q$ and $\MOD_q$ are
$\QAC^0$-equivalent.  By Proposition~\ref{q-reduces-to-2}, $\MOD_q$ is
$\QAC^0$-reducible to $F$.  Hence $F_q$ is $\QAC^0$-reducible to $F$.

Conversely, arrange each block of $\lceil \log q \rceil$ input bits
to an $F_q$ gate as follows. For the control-bit block (which contains
the bit we want to fan out), set all but the last bit to zero, and
call the last bit $b$.  Set all bits in the $i^{th}$ input-bit block
to 0.  Now the $i^{th}$ output of the $F_q$ circuit is $b$,
represented as $\lceil \log q \rceil$ bits with only one possibly
nonzero bit. Send this last output bit $b$ and the input bit $x_i$ to
a controlled-NOT gate.  The outputs of that gate are $b$ and $b \oplus
x_i$.  Now apply $F^{-1}_q$ to the bits that were the outputs of the
$F_q$ gate (which are all left unchanged by the controlled-not's).
This returns all the $b$'s to 0 except for the control bit which is
always unchanged. The outputs of the controlled-not's give the desired
$b \oplus x_i$.  Thus the resulting circuit simulates $F$ with
$\ord(n)$ work bits.
\end{proof}

\begin{theorem}
For any $q \in \nums$, $q \not = 1$, $\QACC = \QACC[q]$.
\end{theorem}
\begin{proof}
  By the preceding lemmas, fanout of bits is equivalent to the $\MOD_q$
function.  Thus we can do fanout, and hence $\MOD_2$, if we can do
$\MOD_q$. By the  result of Proposition~\ref{q-reduces-to-2},
we can do $\MOD_q$ if we can do fanout in constant
depth. Hence $\QACC = \QACC[2] \subseteq \QACC[q]$.
\end{proof}

To compare these results with classical circuits requires a little
care.  For any Boolean function $\phi$ with $n$ inputs and $m$
outputs, we can define a reversible version $\phi'$ on $n+m$ \ bits
where $\phi'(x,y) = (x,y \oplus \phi(x))$ keeps the input $x$ and\
XORs the output $\phi(x)$ with $y$.  Then if $\phi$ has a circuit with
depth $d$ and width $w$, it is easy to construct a reversible circuit
for $\phi'$ of depth $2d-1$ where $wd$ work bits start and end in the
zero state.  We do this by assigning a work bit to each gate in the
original circuit, and replacing each gate with a reversible one that
XORs that work bit with the output.  Then we can erase the work bits
by moving backward through the layers of the circuit.

Then if we adopt the convention that a Boolean function with $n$
inputs and $m$ outputs is in a quantum circuit class if its reversible
version is, we clearly have, for any $k$, $\AC^{k}\ \subseteq
\QACwf^k$ and $\ACC^k \subseteq \QACC^k$.  Thus we have
\[ \AC^0 \subset \ACC^0[2] \subset \ACC^0 \subseteq
      \QACwf^0 = \QACC[q] = \QACC \]
showing that $\QACwf^0$ and $\QACC[2]$ are \
more powerful than
$\AC^0$ and $\ACC^0[2]$ respectively.

Interestingly, if $\QAC^0$ as we first defined it cannot do fanout,
i.e.\ if $\QAC^0 \subset \QACwf^0$, then in a sense it fails to
include $\AC^0$, since the fanout function from $\{0,1\}$ to
$\{0,1\}^n$ is trivially in $\AC^0$.  However, it is not clear whether
it fails to include any $\AC^0$ functions with a one-bit output.  On
the other hand, if $\QAC^0$ can do fanout, it can also do parity and
is greater than $\AC^0$, so either way $\AC^0$ and $\QAC^0$ are
different.  We are indebted to Pascal Tesson for pointing this out.

\section{Upper Bounds}

In this section, we prove the upper bounds results $\NQACCG \subseteq
\TC^0$, $\BQACCG \subseteq \TC^0$, $\NQACCP \subseteq \P/\poly$,
and $\BQACCP \subseteq \P/\poly$.

Suppose $\{F_n\}$ and $\{z_n\}$ determine a language $L$ in
\NQACC. Let $F_n$ be the product of the layers $U_1, \ldots, U_t$ and
$E$ be the distinct entries of the matrices used in the $U_j$'s. By
our definition of $\QACC$, the size of $E$ is fixed with respect to
$n$. We need a canonical way to write sums and products of elements in
$E$ to be able to check $|\bra{\vec{z}}U_1\cdots
U_t\ket{x,0^{p(n)}}|^2>0$ with a $\TC^0$ function. To do this let $A =
\{ \alpha_i\}_{1\leq i\leq m}$ be a maximal algebraically independent
subset of $E$. Let $F=\rats(A)$ and let $B = \{\beta_i\}_{0\leq i <
d}$ be a basis for the field $G$ generated by the elements in $(E -
A)\cup\{1\}$ over $F$.  Since the size of the bases of $F$ and $G$ are
less than the cardinality of $E$ the size of these bases is also fixed
with respect to $n$.

As any sum or product of elements in $E$ is in $G$, it suffices to
come up with a canonical form for elements in $G$. Our representation
is based on Yamakami and Yao~\cite{yy98}. Let $\alpha\in G$. Since $B$
is a basis, $\alpha =\sum^{d-1}_{j=0} \lambda_j \beta_j$ for some
$\lambda_j \in F$. We encode an $\alpha$ as a $d$-tuple (we iterate
the pairing function from the preliminaries to make $d$-tuples)
$\langle \GN{\lambda_0}, \ldots, \GN{\lambda_{d-1}} \rangle$ where
$\GN{\lambda_j}$ encodes $\lambda_j$. As the elements of $A$ are
algebraically independent, each $\lambda_j=s_j/u_j$ where $s_j$ and
$u_j$ are of the form 
\[ \sum_{\vec{k}_j,|\vec{k}_j| \leq e}
a_{\vec{k}_j} (\prod^m_{i=1}\alpha_i^{k_{ij}}).
\]
Here $\vec{k}_j = (k_{1j},\ldots, k_{mj}) \in \ints^m$, $|\vec{k}_j|$
is $\sum_ik_{ij}$, $a_{\vec{k}_j}\in \ints$, and $e \in \nats$. In
particular, any product $\beta_m\cdot\beta_l=\sum^{d-1}_{j=0}
\lambda_j \beta_j$ with $\lambda_j = s_j/u_j$ and $s_j$ and $u_j$ in
this form.  We take a common denominator $u$ for elements of
$E\cup\{\beta_m\cdot\beta_l\}$ and not just $E$ since the
$\lambda_j$'s associated with the $\beta_m\cdot\beta_l$ might have
additional factors in their denominators not in $E$.  Also fix an $e$
large enough to bound the $|\vec{k_j}|$'s which might appear in any
element of $E$ or a product $\beta_m\cdot\beta_l$.  This $e$ will be
constant with respect to $n$. In multiplying $t$ layers of \QACC
\,circuit against an input, the entries in the result will be
polynomial sums and products of elements in
$E\cup\{\beta_m\cdot\beta_l\}$, so we can bound $|\vec{k}_j|$ for
$\vec{k_j}$'s which appear in the $\lambda_j$'s of such an entry by
$e\cdot p(n)$.  To complete our representation of $\alpha\in G$ we
encode $\lambda_j$ as the sequence $\langle r, \langle\langle
a_{\vec{k_j}}, k_{1j}, \ldots, k_{mj} \rangle\rangle\rangle$ where $r$
is the power to which $u$ is raised and $\langle\langle a_{\vec{k_j}},
k_{1j}, \ldots, k_{mj} \rangle\rangle$ is the sequence of $\langle
a_{\vec{k_j}}, k_{1j}, \ldots, k_{mj} \rangle$'s that appear in
$s_j$. By our discussion, the encoding of an $\alpha$ that appears as
an entry in the output after applying a \QACC \,operator to the input
is of polynomial length and so can be manipulated in $\TC^0$.

We have need of the following lemma:

\begin{lem}
\label{sumprod}
Let $p$ be a polynomial.  (1) Let $f(i,x) \in \TC^0$ output encodings
of $a_{i,x}\in\ints[A]$.  Then $\ints[A]$ encodings of
$\sum_{i=1}^{p(|x|)}a_{i,x}$ and $\prod_{i=1}^{p(|x|)}a_{i,x}$ are
$\TC^0$ computable.  (2) Let $f(i,x)\in \TC^0$ output encodings of
$a_{i,x}\in G$. Then $G$ encodings of $\sum_{i=1}^{p(|x|)}a_{i,x}$ and
$\prod_{i=1}^{p(|x|)}a_{i,x}$ are $\TC^0$ computable.
\end{lem}
\begin{proof}
We will abuse notation in this proof and identify the encoding $f(i,x)$
with
its value $a_{i,x}$. So $\sum_i f(i,x)$ and $\prod_i f(i,x)$ will mean the
encoding of $\sum_i a_{i,x}$ and $\prod_i a_{i,x}$ respectively.

(1) To do sums, the first thing we do is form the list $L1=\langle
f(0,x), \ldots, f(p(|x|),x)\rangle$. Then we create a flattened list
$L2$ from this with elements which are the $\langle a_{\vec{k_j}},
k_{1j}, \ldots, k_{mj} \rangle$'s from the $f(i,x)$'s. $L1$ is in
$\TC^0$ using our definition of sequence from the preliminaries, and
closure under sums and $max_i$ to find the length of the longest
$f(i,x)$. To flatten $L1$ we use $max_i$ to find the length $d$ of the
longest $f(i,x)$ for $i\leq p(|x|)$. Then using max twice we can find
the length of the longest $\langle a_{\vec{k_j}}, k_{1j}, \ldots,
k_{mj} \rangle$. This will be the second coordinate in the pair used
to define sequence $L2$. We then do a sum of size $d\cdot p(|x|)$ over
the subentries of $L1$ to get the first coordinate of the pair used to
define $L2$. Given $L2$, we make a list $L3$ of the distinct
$\vec{k_j}$'s that appear as $\langle a_{\vec{k_j}}, k_{1j}, \ldots,
k_{mj} \rangle$ in some $f(i,x)$ for some $i \leq p(|x|)$. This list
can be made from $L2$ using sums, $cond$ and $\mu$. We sum over the $t
\leq length(L2)$ and check if there is some $t'<t$ such that the
$t'$th element of $L2$ has same $\vec{k}_j$ as $t$ and if not add the
$t$th elements $\vec{k_j}$ times 2 raised to the appropriate power. We
know what power by computing the sum of the number of smaller $t'$
that passed this test. Using $cond$ and closure under sums we can
compute in $\TC^0$ a function which takes a list like $L2$ and a
$\vec{k_j}$ and returns the sum of all the $a_{\vec{k_j}}$'s in this
list. So using this function and the lists $L2$ and $L3$ we can
compute the desired encoding.

For products, since the $\alpha_i$'s of $A$ are algebraically
independent, $\ints[A]$ is isomorphic to the polynomial ring
$\ints[y_1,\ldots,y_m]$ under the natural map which takes $\alpha_j$
to $y_j$.  We view our encodings $f(i,x)$ as $m$-variate polynomials
in $\ints[y_1,\ldots,y_m]$. We describe for any $p'$ a circuit that
works for any $\TC^0$ computable $f(i,x)$ such that $\prod_i f(i,x)$
is of degree less than $p'$ viewed as an $m$-variate polynomial. In
$\TC^0$ we define $g(i,x)$ to consist of the sequence of
polynomially many integer values which result from evaluating the
polynomial encoded by $f(i,x)$ at the points $(i_1,\ldots,
i_m)\in\nats^m$ where $0 \leq i_s$ and $\sum_s i_s \leq p'$. To
compute $f(i,x)$ at a point involves computing a polynomial sum of a
polynomial product of integers, and so will be in $\TC^0$. Using
closure under polynomial integer products we compute $k(j,x) :=\prod_i
\beta(j, g(i,x))$ where $\beta$ is the sequence projection function
from the preliminaries. Our choice of points is what is called by
Chung and Yao~\cite{chungyao77} the {\em $p'$-th order principal
lattice} of the $m$-simplex given by the origin and the points $p'$
from the origin in each coordinate axis.  By Theorems~1 and 4 of that
paper (proved earlier by a harder argument in Nicolaides~\cite{nic72})
the multivariate Lagrange Interpolant of degree $p'$ through the
points $k(j,x)$ is unique. This interpolant is of the form
$P(y_1,\ldots,y_m) = \sum_j p_j(y_1,\ldots,y_m)k(j,x)$ where the
$p_j$'s are polynomials which do not depend on the function $f$. An
explicit formula for these $p_j$'s is given in Corollary~2 of Chung
and Yao~\cite{chungyao77} as a polynomial product of linear
factors. Since these polynomials are all of degree less than $p'$,
they have only polynomial in $p'$ many coefficients and in PTIME these
coefficients can be computed by iteratively multiplying the linear
factors together. We can then hard code these $p_j$'s (since they
don't depend on $f$) into our circuit and with these $p_j$'s,
$k(j,x)$, and closure under sums we can compute the polynomial of the
desired product in $\TC^0$.

(2) We do sums first. Assume $f(i,x)
:=\sum^{d-1}_{j=0}\lambda_{ij}\beta_j$.  One immediate problem is that
the $\lambda_{ij}$ and $\lambda_{i'j}$ might use different $u^r$'s for
their denominators.  Since $\TC^0$ is closed under poly-sized maximum,
it can find the maximum value $r_0$ to which $u$ is raised. Then it
can define a function $g(i,x) = \sum^{d-1}_{j=0}\gamma_{ij}\beta_j$
which encodes the same element of $G$ as $f(i,x)$ but where the
denominators of the $\gamma_{ij}$'s are now $u^{r_0}$. If $\lambda_j$
was $s_j/u^r$ we need to compute the encoding $s_j\cdot
u^{r_0-r}/u^{r_0}$. This is straightforward from (1). Now
\[ \sum_{i=1}^{p(|x|)}f(i,x) = \sum_{i=1}^{p(|x|)}g(i,x) =
\sum^{d-1}_{j=0}[(\sum_{i=1}^{p(|x|)}s_{ij})/u^{r_0}]\beta_j,
\]
where $s_{ij}$'s are the numerators of the $\gamma_{ij}$'s in
$g(i,x)$.  From part (1) we can compute the encoding $e_j$ of
$(\sum_{i=1}^{p(|x|)}s_{ij})$ in $\TC^0$.  So the desired answer
$\langle \langle r_0, e_0\rangle, \cdots, \langle r_0, e_{d-1}\rangle
\rangle$ is in $\TC^0$.

For products $\prod_{i=1}^{p(|x|)}f(i,x)$, we play the same trick as
the in the $\ints[A]$ product case. We view our encodings of elements
of $G$ as d-variate polynomials in $F(y_0,\ldots,y_{d-1})$ under the
map $\beta_k$ goes to $y_k$. (Note that this map is not necessarily an
isomorphism.) We then create a function $g(i,x)$ which consists of the
sequence of values obtained by evaluating $f(i,x)$ at polynomially
many points in a lattice as in the first part of this
lemma. Evaluating $f(i,x)$ at a point can easily be done using the
first part of this lemma. We then use part (1) of this lemma to
compute the products $k(j,x) = \beta(j,g(i,x))$. We then get the
interpolant $P(y_0, \ldots, y_{d-1}) = \sum_j
p_j(y_0,\ldots,y_m)k(j,x)$.  We non-uniformly obtain the encoding of
$p_j(\beta_0,\ldots, \beta_{d-1})$ expressed as an element of
$G$. i.e., in the form $\sum^{d-1}_{w=0} \lambda_{jw}\beta_w$.  Thus,
the product $\prod_{i=1}^{p(|x|)}f(i,x)$ is
\[ \sum^{d-1}_{w=0}(\sum_j\lambda_{jw}k(j,w)) \beta_w \]
The encoding of the products is the d-tuple given by $\langle
\sum_j\lambda_{j0}k(j,0),\ldots,\sum_j\lambda_{jd-1}k(j,d-1) \rangle$.
Each of its components is a polynomial sum of a product of two things
in $F$ and can be computed using the first part of the lemma.
\end{proof}

For $\{F_n\} \in \QAC^0_{wf}= \QACC$, the vectors that $F_n$ act on
are elements of a $2^{n+p(n)}$ dimensional space ${\mathcal
E}_{1,n+p(n)}$ which is a tensor product of the 2-dimensional spaces
${\mathcal E}_1, \ldots {\mathcal E}_{n+p(n)}$, which in turn are each
spanned by $\ket{0}, \ket{1}$. We write ${\mathcal E}_{j,k}$ for the
subspace $\otimes^{k}_{i=j} {\mathcal E}_i$ of ${\mathcal
E}_{1,n+p(n)}$.  We now define a succinct way to represent a set of
vectors in ${\mathcal E}_{1,n+p(n)}$ which is useful in our argument
below. A {\em tensor graph} is a directed acyclic graph with one
source node of indegree zero, one terminal node of outdegree zero, and
two kinds of edges: horizontal edges, which are unlabeled, and
vertical edges, which are labeled with a pair of amplitudes and a
product of {\em colors} and {\em anticolors} (which are defined below). 
We require that all paths from the source to the
terminal traverse the same number of vertical edges and that no vertex
can have vertical edge indegree greater than one or outdegree greater
than one. The {\em height} of a node in a tensor graph is the number
of vertical edges traversed to get to it on any path from the source;
the {\em height} of an edge is the height of its end node.  The {\em
width} of a tensor graph is maximum number of nodes of the same
height. As an example of a tensor graph where our color product is the
number 1, consider the following figure:

\begin{center}
\begin{picture}(2025,2343)(2101,-2707)
\thicklines
\put(3226,-511){\circle{150}}
\put(3226,-1186){\circle{150}}
\put(3226,-1861){\circle{150}}
\put(3226,-2536){\circle{150}}
\put(3226,-586){\line( 0,-1){525}}
\put(3226,-1261){\line( 0,-1){525}}
\put(3226,-1936){\line( 0,-1){525}}
\put(3977,-511){\circle{150}}
\put(3977,-1186){\circle{150}}
\put(3977,-1861){\circle{150}}
\put(3977,-2536){\circle{150}}
\put(3977,-586){\line( 0,-1){525}}
\put(3977,-1261){\line( 0,-1){525}}
\put(3977,-1936){\line( 0,-1){525}}
\put(3301,-2536){\line( 1, 0){600}}
\put(3301,-511){\line( 1, 0){600}}
\multiput(3151,-1036)(6.25000,-6.25000){13}{\makebox(6.6667,10.0000){
\SetFigFont{7}{8.4}{rm}.}}
\multiput(3226,-1111)(6.25000,6.25000){13}{\makebox(6.6667,10.0000){
\SetFigFont{7}{8.4}{rm}.}}
\multiput(3826,-436)(6.25000,-6.25000){13}{\makebox(6.6667,10.0000){
\SetFigFont{7}{8.4}{rm}.}}
\multiput(3901,-511)(-6.25000,-6.25000){13}{\makebox(6.6667,10.0000){
\SetFigFont{7}{8.4}{rm}.}}
\multiput(3901,-1036)(6.25000,-6.25000){13}{\makebox(6.6667,10.0000){
\SetFigFont{7}{8.4}{rm}.}}
\multiput(3976,-1111)(6.25000,6.25000){13}{\makebox(6.6667,10.0000){
\SetFigFont{7}{8.4}{rm}.}}
\multiput(3151,-1711)(6.25000,-6.25000){13}{\makebox(6.6667,10.0000){
\SetFigFont{7}{8.4}{rm}.}}
\multiput(3226,-1786)(6.25000,6.25000){13}{\makebox(6.6667,10.0000){
\SetFigFont{7}{8.4}{rm}.}}
\multiput(3151,-2386)(6.25000,-6.25000){13}{\makebox(6.6667,10.0000){
\SetFigFont{7}{8.4}{rm}.}}
\multiput(3226,-2461)(6.25000,6.25000){13}{\makebox(6.6667,10.0000){
\SetFigFont{7}{8.4}{rm}.}}
\multiput(3901,-1711)(6.25000,-6.25000){13}{\makebox(6.6667,10.0000){
\SetFigFont{7}{8.4}{rm}.}}
\multiput(3976,-1786)(6.25000,6.25000){13}{\makebox(6.6667,10.0000){
\SetFigFont{7}{8.4}{rm}.}}
\multiput(3901,-2386)(6.25000,-6.25000){13}{\makebox(6.6667,10.0000){
\SetFigFont{7}{8.4}{rm}.}}
\multiput(3976,-2461)(6.25000,6.25000){13}{\makebox(6.6667,10.0000){
\SetFigFont{7}{8.4}{rm}.}}
\multiput(3351,-2611)(-6.25000,6.25000){13}{\makebox(6.6667,10.0000){
\SetFigFont{7}{8.4}{rm}.}}
\multiput(3276,-2536)(6.25000,6.25000){13}{\makebox(6.6667,10.0000){
\SetFigFont{7}{8.4}{rm}.}}
\put(3076,-436){\makebox(0,0)[lb]{\smash{\SetFigFont{12}{14.4}{rm}s}}}
\put(3076,-2686){\makebox(0,0)[lb]{\smash{\SetFigFont{12}{14.4}{rm}t}}}
\put(2326,-886){\makebox(0,0)[lb]{\smash{\SetFigFont{10}{16.8}{rm}\{1\} 
0,1}}}
\put(2101,-1561){\makebox(0,0)[lb]{\smash{\SetFigFont{10}{16.8}{rm}$\{1\}
 \frac{1}{\sqrt{2}},\frac{1}{\sqrt{2}}$}}}
\put(2101,-2236){\makebox(0,0)[lb]{\smash{
\SetFigFont{10}{16.8}{rm}\{1\} 1/2,0}}}
\put(4126,-886){\makebox(0,0)[lb]{\smash{\SetFigFont{10}{16.8}{rm}\{1\} 
1,0}}}
\put(4126,-1561){\makebox(0,0)[lb]{\smash{\SetFigFont{10}{16.8}{rm}$\{1\}
\frac{1}{\sqrt{2}},\frac{-1}{\sqrt{2}}$}}}
\put(4051,-2236){\makebox(0,0)[lb]{\smash{
\SetFigFont{10}{16.8}{rm}\{1\} 1/2,0}}}
\end{picture}
\end{center}
The rough idea of tensor graphs is that paths through the graph
correspond to collections of vectors in ${\mathcal E}_{1,n}$. For this
particular figure the left path from the source node (s) to the
terminal node (t) corresponds to the vectors given by
\[ \ket{1}\otimes (\frac{1}{\sqrt{2}}\ket{0} +
\frac{1}{\sqrt{2}}\ket{1})\otimes \frac{1}{2}\ket{0} \]
and the right hand path corresponds to
\[ \ket{0}\otimes (\frac{1}{\sqrt{2}}\ket{0} +
\frac{-1}{\sqrt{2}}\ket{1})\otimes \frac{1}{2}\ket{0}. \]

A ${\mathcal E}_{j,k}$-{\em term} in a tensor graph is a maximal
induced tensor subgraph between a node of height $j-1$ and a node of
height $k$. If the horizontal indegree of the node at height $j-1$ is
zero and the horizontal outdegree of the node at height $k$ is zero
then we say the term is {\em good}.  For the graph we considered above
there are two good ${\mathcal E}_{1,2}$-terms
 and two good
${\mathcal E}_{2,3}$-terms but only one ${\mathcal
E}_{1,3}$-term
 corresponding to the whole figure.

``Colors" are used to handle controlled-not layers. A color $c$ and its
anticolor $\tilde{c}$ are defined to obey the following multiplicative
properties:
$c\cdot c =\tilde{c} \cdot \tilde{c} = 1$ and $c \cdot \tilde{c} = 0$.
Given a color $b$ and a product of colors $c$ not involving $b$ or its
anticolor we require $b \cdot c = c \cdot b$ and
 $\tilde{b} \cdot c = c
\cdot \tilde{b}$. If $a$ is a product of colors and
 anticolors not
involving the color $b$ or $\tilde{b}$ and $c$
 is another product of
colors we have $a(bc)=(ab)c$. 
 We consider formal sums of products
of complex numbers times colors.
 We require complex numbers to
commute with colors and require colors and
 anticolors to distribute,
i.e., if $a$, $b$, $c$ are colors or anticolors then $a\cdot (b+c) =
a\cdot b + a\cdot c$ and $(b+c)\cdot a = b\cdot a + c \cdot
a$. Finally, we require addition to work so that the above structure
satisfies the axioms of an $\complexes$-algebra.  Given a tensor graph
$G$ denote this $\complexes$-algebra by $\mathcal{A}_{G}$.  Since
\[ (a\cdot a)\cdot \tilde{a} = \tilde{a} \neq 0=a\cdot(a \cdot\tilde{a}) 
\] 
this algebra is not associative. However, in the sums we will consider,
the terms will never have more than two positions where a color or its
anticolor can occur, so the products we will consider are associative.

Using our our earlier encoding for the elements of $\complexes$ which
could appear in a $\QACC$ computation, it is straightforward to use
sequence coding to get a $\TC^0$ encodings of the relevant elements of
$\mathcal{A}_G$. As an example of how colors affect amplitudes,
consider the following picture:

\begin{center}
\begin{picture}(2133,2466)(2101,-2719)
\thicklines
\put(3226,-511){\circle{150}}
\put(3226,-1186){\circle{150}}
\put(3226,-1861){\circle{150}}
\put(3226,-2536){\circle{150}}
\put(3226,-586){\line( 0,-1){525}}
\put(3226,-1261){\line( 0,-1){525}}
\put(3226,-1936){\line( 0,-1){525}}
\put(3977,-511){\circle{150}}
\put(3977,-1186){\circle{150}}
\put(3977,-1861){\circle{150}}
\put(3977,-2536){\circle{150}}
\put(3376,-361){\circle{20}}
\put(3601,-286){\circle{20}}
\put(3826,-361){\circle{20}}
\put(4126,-661){\circle{20}}
\put(4201,-811){\circle{20}}
\put(4126,-1036){\circle{20}}
\put(3451,-1336){\circle{20}}
\put(3826,-1336){\circle{20}}
\put(3656,-1411){\circle{20}}
\put(3451,-1486){\circle{20}}
\put(3376,-1711){\circle{20}}
\put(3451,-1636){\circle{20}}
\put(3076,-661){\circle{20}}
\put(3001,-811){\circle{20}}
\put(3001,-1036){\circle{20}}
\put(3001,-1261){\circle{20}}
\put(3001,-1486){\circle{20}}
\put(3001,-1711){\circle{20}}
\put(3001,-1936){\circle{20}}
\put(3076,-2236){\circle{20}}
\put(3751,-1711){\circle{20}}
\put(3601,-1711){\circle{20}}
\put(4126,-2011){\circle{20}}
\put(4201,-2236){\circle{20}}
\put(4126,-2386){\circle{20}}
\put(3451,-2686){\circle{20}}
\put(3676,-2686){\circle{20}}
\put(3901,-2611){\circle{20}}
\put(3301,-2536){\line( 1, 0){600}}
\put(3301,-511){\line( 1, 0){600}}
\multiput(3151,-1036)(6.25000,-6.25000){13}{\makebox(6.6667,10.0000)
{\SetFigFont{7}{8.4}{rm}.}}
\multiput(3226,-1111)(6.25000,6.25000){13}{\makebox(6.6667,10.0000)
{\SetFigFont{7}{8.4}{rm}.}}
\multiput(3826,-436)(6.25000,-6.25000){13}{\makebox(6.6667,10.0000)
{\SetFigFont{7}{8.4}{rm}.}}
\multiput(3901,-511)(-6.25000,-6.25000){13}{\makebox(6.6667,10.0000)
{\SetFigFont{7}{8.4}{rm}.}}
\multiput(3901,-1036)(6.25000,-6.25000){13}{\makebox(6.6667,10.0000)
{\SetFigFont{7}{8.4}{rm}.}}
\multiput(3976,-1111)(6.25000,6.25000){13}{\makebox(6.6667,10.0000)
{\SetFigFont{7}{8.4}{rm}.}}
\multiput(3151,-1711)(6.25000,-6.25000){13}{\makebox(6.6667,10.0000)
{\SetFigFont{7}{8.4}{rm}.}}
\multiput(3226,-1786)(6.25000,6.25000){13}{\makebox(6.6667,10.0000)
{\SetFigFont{7}{8.4}{rm}.}}
\multiput(3151,-2386)(6.25000,-6.25000){13}{\makebox(6.6667,10.0000)
{\SetFigFont{7}{8.4}{rm}.}}
\multiput(3226,-2461)(6.25000,6.25000){13}{\makebox(6.6667,10.0000)
{\SetFigFont{7}{8.4}{rm}.}}
\multiput(3901,-2386)(6.25000,-6.25000){13}{\makebox(6.6667,10.0000)
{\SetFigFont{7}{8.4}{rm}.}}
\multiput(3976,-2461)(6.25000,6.25000){13}{\makebox(6.6667,10.0000)
{\SetFigFont{7}{8.4}{rm}.}}
\multiput(3351,-2611)(-6.25000,6.25000){13}{\makebox(6.6667,10.0000){
\SetFigFont{7}{8.4}{rm}.}}
\multiput(3276,-2536)(6.25000,6.25000){13}{\makebox(6.6667,10.0000){
\SetFigFont{7}{8.4}{rm}.}}
\multiput(3376,-1261)(-6.25000,6.25000){13}{\makebox(6.6667,10.0000){
\SetFigFont{7}{8.4}{rm}.}}
\multiput(3301,-1186)(6.25000,6.25000){13}{\makebox(6.6667,10.0000){
\SetFigFont{7}{8.4}{rm}.}}
\multiput(3826,-1786)(6.25000,-6.25000){13}{\makebox(6.6667,10.0000){
\SetFigFont{7}{8.4}{rm}.}}
\multiput(3901,-1861)(-6.25000,-6.25000){13}{\makebox(6.6667,10.0000){
\SetFigFont{7}{8.4}{rm}.}}
\put(3977,-586){\line( 0,-1){525}}
\put(3977,-1936){\line( 0,-1){525}}
\put(3901,-1186){\line(-1, 0){600}}
\put(3901,-1861){\line(-1, 0){600}}
\put(3301,-2611){\line( 0, 1){  0}}
\put(3301,-2611){\line( 1, 0){ 75}}
\put(3301,-2686){\line( 0, 1){ 75}}
\put(4051,-2386){\line( 0,-1){ 75}}
\put(4051,-2461){\line( 1, 0){ 75}}
\put(3826,-1786){\line( 1, 0){ 75}}
\put(3901,-1786){\line( 0, 1){ 75}}
\put(3826,-436){\line( 1, 0){ 75}}
\put(3901,-436){\line( 0, 1){ 75}}
\put(4051,-1036){\line( 0,-1){ 75}}
\put(4051,-1111){\line( 1, 0){ 75}}
\put(3301,-1261){\line( 0, 1){  0}}
\put(3301,-1261){\line( 1, 0){ 75}}
\put(3301,-1261){\line( 0,-1){ 75}}
\put(3301,-1711){\line( 0,-1){ 75}}
\put(3301,-1786){\line( 1, 0){ 75}}
\put(3151,-2311){\line( 0,-1){ 75}}
\put(3151,-2386){\line(-1, 0){ 75}}
\put(3076,-436){\makebox(0,0)[lb]{\smash{\SetFigFont{12}{14.4}{rm}s}}}
\put(3076,-2686){\makebox(0,0)[lb]{\smash{\SetFigFont{12}{14.4}{rm}t}}}
\put(1931,-886){\makebox(0,0)[lb]{\smash{
\SetFigFont{10}{16.8}{rm}\{$b$\}$\frac{-1}{\sqrt{2}},\frac{-1}{\sqrt{2}}$}}}
\put(1931,-1561){\makebox(0,0)[lb]{\smash{
\SetFigFont{10}{16.8}{rm}\{1\}$\frac{-1}{\sqrt{2}},\frac{1}{\sqrt{2}}$}}}
\put(2051,-2236){\makebox(0,0)[lb]{\smash{
\SetFigFont{10}{16.8}{rm}\{$b$\} 1, 0}}}
\put(4200,-886){\makebox(0,0)[lb]{\smash{
\SetFigFont{10}{16.8}{rm}\{$\tilde{b}$\}
$\frac{1}{\sqrt{2}},\frac{-1}{\sqrt{2}}$}}}
\put(4200,-2236){\makebox(0,0)[lb]{\smash{
\SetFigFont{10}{16.8}{rm}\{$\tilde{b}$\} 0,1}}}
\end{picture}
\end{center}
The amplitude of $\ket{1,0,0}$ in the left hand dotted path is $b
\cdot \frac{-1}{\sqrt{2}} \cdot 1 \cdot \frac{-1}{\sqrt{2}} \cdot b
\cdot 1 = 1/2$ using commutativity and $b^2=1$. Its amplitude in the
right hand dotted path would be zero because of the last vertical
edge. However, vectors such as $\ket{0,0,1}$ would have nonzero
amplitude in the right hand dotted path.  Nevertheless, the amplitude
of any vector $\ket{\vec{x}}$ in any path other than the dotted ones
from $s$ to $t$ will be $0$ as $b\cdot\tilde{b}=0$. More formally, we
define the amplitude of an $\ket{x}$ in a vertical edge as equal to
the left amplitude times the color product in the edge if $\ket{x}$ is
$\ket{0}$ and equal to the right amplitude times the color product in
the edge if $\ket{x}$ is $\ket{1}$.  The amplitude of a vector
$\ket{x_1,\ldots,x_j}$ in a path in a tensor graph is the product over
$k$ from 1 to $j$ of the amplitude of the vectors $\ket{x_k}$ in the
vertical edge of height $k$. The amplitude of a vector
$\ket{x_j,\ldots, x_k}$ in an ${\mathcal E}_{j,k}$-term is the sum of
its amplitude in its paths. The amplitude of a vector
$\ket{x_1,\ldots,x_{p(n)}}$ in a tensor graph $G$ is defined to be the
sum of its amplitudes in $G$'s ${\mathcal E}_{1,p(n)}$-terms.

As we will be interested in families of tensor graphs $\{G_n\}$,
corresponding to our circuit families we want to look at those
families with a certain degree of uniformity. We say a family of
tensor graphs $\{G_n\}$ is {\em color consistent} if: (1) the number
of colors for edges of the same height is bounded by a constant $k$
with respect to $n$, (2) the number of heights in which a given
color/anticolor can appear is exactly two (colors and their anticolors
must appear on the same heights), (3) each color product at the same
height is of the form $\prod^k_{i=0} l_i$ where $l_i$ must be either a
color $c_i$ or $\tilde{c_i}$ (it follows there are $2^k$ possible
color products for edges at a given height). We say that a
color/anticolor is {\em active} at a given height if the height is at
or after the first height at which the color/anticolor occurs and is
below the height of its second occurrence. The family is further said
to be {\em log-color depth } if the number of active colors/anticolors
of a given height is log-bounded.
\begin{thm}
\label{multlayers}
Let $\{F_n\}$ be a family of \QACC\, operators and let
$\{\bra{\vec{z}_n}\}$ a family of observables. (1) There is a
color-consistent family of tensor graphs of width $2^{2^{2t}}$ and
polynomial size representing the output amplitudes of $U_1\cdots U_t
\ket{\vec{z}_n}$ where $U_i$ are the layers of $F_n$.  (2) If
$\{F_n\}$ is in $\QACCP$ then the family of tensor graphs will be of
log-color depth. (3) If $\{F_n\}$ is in $\QACCG$ then the number of
paths from the source to the terminal node is polynomially bounded.
\end{thm} 
\begin{proof}
The proof is by induction on $t$. In the base case, $t=0$, we do not
multiply any layers, and we can easily represent this as a tensor
graph of width 1. Assume for $j <t$ that $U_j\cdots U_1\ket{\vec{x},
0^{p(n)}}$ can be written as color consistent tensor graph of width
$2^{2^{2t}}$ and polynomial size. There are two cases to consider: In
the first case the layer is a tensor product of matrices $M_1 \otimes
\cdots \otimes M_\nu$ where the $M_k$'s are Toffoli gates, one qubit
gates, or fan-out gates (since $\QAC^0_{wf} = \QACC$); in the second
case the layer is a controlled-not layer.

For the first case we ``multiply'' $U_t$ against our current graph by 
``multiplying'' each $M_j$ in parallel against the terms in our sum 
corresponding to $M_j$'s domain, say ${\mathcal E}_{j',k'}$. If  $M_j =
\left(
\begin{array}{cc} u_{00} & u_{01} \\ u_{10} &  u_{11}\end{array} \right)$
with
domain ${\mathcal E}_{j'}$ is a one-qubit gate, then we multiply the
two amplitudes in each vertical edge of height $j'$ in our tensor
graph by $M_j$.  This does not effect the width, size, or number of
paths through the graph.  If $M_j$ is a Toffoli gate, then for each
good term $S$ in ${\mathcal E}_{j',k'}$ in
 our tensor graph we add
one new term to the resulting graph. This term is added by adding a
horizontal edge going out from the source node of $S$ followed by the
new ${\mathcal E}_{j',k'}$-term followed by a horizontal edge into the
terminal node of $S$. The new term is obtained from $S$ by setting
to
$0$ the left hand amplitudes of all edges in $S$ of height between
$j'$ and $k'-1$ and then if $\alpha,\gamma$ is the amplitude of an
edge of height $k'$ in the new term we change it to $\gamma -\alpha,
\alpha-\gamma$. This new term adjusts the amplitude for the case of a
$\ket{1}^{\otimes(k'-j'-1)}$ vector in ${\mathcal E}_{j',k'-1}$
tensored with either a $\ket{0}$ or $\ket{1}$.  This operation
increases the width of the new tensor graph by the width of the good
${\mathcal E}_{j',k'}$-term for each good ${\mathcal E}_{j',k'}$-term
in the graph. Since the original graph has
 width $2^{2^{2(t-1)}}$
there are at most this many starting and ending vertices for such
terms.  So there at most $(2^{2^{2(t-1)}})^2$ such terms.  Each of
these terms has width at most $2^{2^{2(t-1)}}$.  Thus, the new width
is at most
\[ 2^{2^{2(t-1)}} + (2^{2^{2(t-1)}})^2\cdot 2^{2^{2(t-1)}}< 2^{2^{2t}}. \]
Notice this action adds one new path through the ${\mathcal
E}_{j',k'}$ part of the graph for every existing one.

Now suppose $M_j$ is a fan-out gate, let $S$ be a good ${\mathcal
E}_{j',k'}$-term in our tensor graph
 and let $e$ be any vertical
edge in $S$ in ${\mathcal E}_{k'}$.  Suppose $e$
 has amplitude
$\alpha$ for $\ket{0}$ and amplitude $\gamma$ for $\ket{1}$. In
 the
new graph we change the amplitude of $e$ to $\alpha,0$. We then add
a
 horizontal edge out of the source node of $S$ followed by a new
${\mathcal E}_{j',k'}$-term followed by a horizontal edge into the
terminal node of $S$. The new term is obtained from $S$ by changing
the amplitude for edges in ${\mathcal E}_{k'}$ with amplitudes
$\alpha,\gamma$ in $S$ to $0,\gamma$.  The amplitudes of the
non-${\mathcal E}_{k'}$ edges in this term are the reverse of the
corresponding edge in $S$, i.e., if the edge in $S$ had amplitude
$\delta,\zeta$ then the new term edge would have amplitude
$\zeta,\delta$. The same argument as in the Toffoli case shows the new
width is bounded by $2^{2^{2t}}$ and that this action adds one new
path through the ${\mathcal E}_{j',k'}$ part of the graph for every
existing one.

For the case of a controlled-not layer, suppose we have a
controlled-not going from line $i$ onto line $j$.  Let $c, \bar{c}$ be
a new color, anti-color pair not yet appearing in the graph. Let $e_i$
be a vertical edge of height $i$ in the graph and let $C_i,
\alpha_i,\gamma_i$ be respectively its color product and two
amplitudes.  Similarly, let $e_j$ be a vertical edge of height $j$ in
the graph and $C_j, \alpha_j,\gamma_j$ be its color product and two
amplitudes. In the new graph we multiply $c$ times the color product
of $e_i$ and $e_j$ and change the amplitude of $e_i$ to
$\alpha_i,0$. We then add a horizontal edge going out from the
starting node of $e_i$, followed by a vertical edge with values
$C_i\cdot\tilde{c}, 0, \gamma_i$ followed by a horizontal edge into
the terminal node of $e_i$. In turn, we add a horizontal edge going
out of the starting node of $e_j$, followed by a vertical edge with
values $C_j\cdot\tilde{c}, \gamma_i, \alpha_j$ followed by a
horizontal edge into the terminal node of $e_j$.  We handle all other
controlled gates in this layer in a similar fashion (recall they must
go to disjoint lines).  We add at most a new vertex of a given height
for every existing vertex of a given height. So the total width is at
most doubled by this operation and $2\cdot2^{2^{2(t-1)}} <
2^{2^{2t}}$.  In the $\QACCP$ case, simulating a layer which is a
Kronecker product 
 of spaced controlled-not gates and identity
matrices, notice we would at most
 add one to the color depth at any
place. So if a controlled-not layer is
 a composition of $\ord(\log)$
many such layers it will increase the color depth
 by
$\ord(\log)$. In the $\QACCG$ case, notice that simulating a single
controlled-not
 we add one new path for each existing path through
the graph at each of
 the two heights affected. This gives three new
paths on the whole subspace
 for each old one.

Since we have handled the two possible layer cases and the changes we
needed
to make only increase the resulting tensor graph polynomially, we
thus have established the induction step and (1) and (2) of the theorem.
For (3), observe for each multi-line gate we handle in adding a layer
we at most quadruple the number of paths through the subspace where that
gate
applies. Since there are at most logarithmically many such gates, the
number
of paths through the graph increases polynomially.
\end{proof}

\begin{theorem}
\label{amplitudes}
Let $\{G_n\}$ be a family of constant width color-consistent tensor
graphs of vectors in ${\mathcal E}_{1,p(n)}$. Assume the
coefficients of amplitudes in the $\{G_n\}$ can be encoded in
$\TC^0$ using our encoding scheme described earlier and that
$\{G_n\}$ has log-color depth.  Then the amplitude of any basis vector
of ${\mathcal E}_{1,p(n)}$ in $G_n$ is P/poly computable. If the
number of paths through the graph from the source to the terminal node
is polynomially bounded then the amplitude of any basis vector is
$\TC^0$ computable.

\end{theorem}
\begin{proof}
Let $G_n$ be a particular graph in the family and let
$\ket{\vec{x}_n}$ be the vector whose amplitude we want to compute.
Assume that all graphs in our family have fewer than $k$ colors in any
color product and have a width bounded by $w$.  We will proceed from
the source to the terminal node one height at a time to compute the
amplitude. Since the width is $w$ the number of ${\mathcal E}_1$-terms
is at most $w$ and each of these must have width at most $w$. Let
$\alpha_{1,1}, \ldots, \alpha_{1,w}$ (some of which may be zero)
denote the amplitudes in ${\mathcal A}_{G_n}$ of $\ket{x_{n,1}}$ in
each of these terms. The $\alpha_{1,i}$ are each sums of at most $w$
amplitudes times the color products of at most $k$ colors and
anticolors, so the encoding of these $w$ amplitudes is $\TC^0$
computable. Because of the restriction on the width of $G_n$ there are
at most $w$ many ${\mathcal E}_{1,j}$-terms, $w^2$ many ${\mathcal
E}_{j,j+1}$-terms, and $w$ many ${\mathcal E}_{1,j+1}$-terms. Fixing
some ordering on the nodes of height $j$ and $j+1$ let
$\gamma_{j,i,k}$ be the amplitude of $\ket{x_{n,j+1}}$ in the
${\mathcal E}_{j,j+1}$-term with source the $i$th node of height $j$
and with terminal node the $k$th node of height $j+1$. The amplitude
is zero if there is no such ${\mathcal E}_{j,j+1}$-term. Then the
amplitudes $\alpha_{j+1,1}, \ldots, \alpha_{j+1,w}$ of the ${\mathcal
E}_{1,j+1}$-terms can be computed from the amplitudes $\alpha_{j,1},
\ldots, \alpha_{j,w}$ of the ${\mathcal E}_{1,j}$-terms using the
formula
\[ \alpha_{j+1,k} = \sum_{i=1}^w \alpha_{j,i}\cdot \gamma_{j,i,k}. \] 
Thus $\alpha_{j+1,k}$ can be computed from the $\alpha_{j,i}$ using
a polynomial sized circuit to do these adds and multiplies. Similarly,
each $\alpha_{j,k}$ can be computed by polynomial sized circuits from the
$\alpha_{j-1,k}$'s and so on. Since we have log-color depth the number of
terms consisting of elements in our field times color products in a
$\alpha_{j,k}$ will be polynomial. So the size of the $\alpha_{j,k}$'s $j
\leq
p(n)$, $k \leq w$ will be polynomial in the input $\vec{x}_n$. So the size
of
the circuits for each $\alpha_{j,k}$ where $j\leq p(n)$ and $k\leq w$ will
be
polynomial size. There is only one ${\mathcal E}_{1,p(n)}$-term in $G_n$
and
its amplitude is that of $\ket{\vec{x}_n}$, so this shows it has polynomial
sized circuits.  
For the $\TC^0$ result, if the number of paths is polynomially
bounded, then the amplitude can be written as the polynomial sum of
the amplitudes in each path. The amplitude in a path can in turn be
calculated as a polynomial product of the amplitudes times the colors
on the vertical edges in the path.  Our condition on every color
appearing at exactly two heights guarantees the color product along
the whole path will be 1 or 0, and will be zero iff we get a color and
its anticolor on the path. This is straightforward to check in
$\TC^0$, so this sum of products can thus be computed in $\TC^0$
using Lemma~\ref{sumprod}.
\end{proof}

\begin{cor}
\item[ (1)] $\EQACCP \subseteq \NQACCP \subseteq \P/\poly$, and  
$\BQACCP \subseteq \P/\poly$.
\item[ (2)] $\EQACCG \subseteq \NQACCG \subseteq \TC^0$, and $\BQACCG
\subseteq \TC^0$.
 
\end{cor}
\begin{proof}
Given a a family $\{F_n\}$ of $\QACCP$ operators and a family
$\{\bra{\vec{z}_n}\}$ of states we can use Theorem~\ref{multlayers} to
get a family $\{G_n\}$ of log color depth, color-consistent tensor
graphs representing the amplitudes of $F^{-1}_n\ket{\vec{z}_n}$. Note
$\{F^{-1}_n\}$ is also a family of $\QACCP$ operators since Toffoli
and fan-out gates are their
 own inverses, the inverse of any one
qubit gate is also a one qubit gate
 (albeit usually a different
one), and finally a controlled-not layer is its own
inverse. Theorem~\ref{amplitudes} shows there is a P/poly circuit
computing the amplitude of any vector $\ket{\vec{x}_n}$ in this
graph. This amounts to calculating
\[ \bra{\vec{x}_n} F^{-1}_n \ket{\vec{z}_n} = \bra{\vec{z}_n} F_n
\ket{\vec{x}_n} . 
\]
If this is nonzero, then $|\bra{\vec{z}_n} F_n \ket{\vec{x}_n}|^2 >0$,
and we know $\vec{x}$ is in the language. In the $\BQACC_{\rats}$ case
everything is a rational so P/poly can explicitly compute the
magnitude of the amplitude and check if it is greater than $3/4$.  The
$\TC^0$ result follows similarly from the $\TC^0$ part of
Theorem~\ref{multlayers}.
\end{proof}

Finally, we note that some of the inclusions in  the previous corollary
can be strengthened if we assume that the circuit families are
polynomial-time uniform and their coefficients polynomial-time
computable. In particular
p-uniform $\NQACCP$ is contained  in P
and p-uniform $\NQACCG$ is contained in p-uniform $\TC^0$.

\section{Discussion and Open Problems}

A number of open questions are suggested by our work.

\begin{itemize}
\item Is $\QAC^0 = \QACwf^0$?  That is, can the fanout gate be
constructed in constant depth when each qubit can only act as an
input to one gate in each layer?
\item Is $\QACwf^0 = \QTC^0$?  That is, can the techniques used here
be extended to construct quantum threshold gates in constant depth?
\item Is all of $\NQACC$ in $\TC^0$ or even $\P/\poly$?  We
conjecture that $\NQACC$ is in $\TC^0$. As mentioned in the
introduction, we have developed techniques that remove some of the
important obstacles to proving this.
\item Are there any natural problems in $\NQACC$ that are not known to
be in $\ACC$?
\item What exactly is the complexity of the languages in $\EQACC$,
$\NQACC$ and $\BQACC_{\rats}$?  We entertain two extreme
possibilities.  Recall that the class $\ACC$ can be computed by
quasipolynomial size depth 3 threshold circuits~\cite{yao90}. It would
be quite remarkable if $\EQACC$ could also be simulated in that
manner. However, it is far from clear if any of the techniques used in
the simulations of $\ACC$ (the Valiant-Vazirani lemma, composition of
low-degree polynomials, modulus amplification via the Toda
polynomials, etc.), which seem to be inherently irreversible, can be
applied in the quantum setting.  At the other extreme, it would be
equally remarkable if $\NQACC$ and $\NQTC^0$ (or $\BQACC_{\rats}$ and
$\NQTC^0$) coincide.  Unfortunately, an optimal characterization of
$\QACC$ language classes anywhere between those two extremes would
probably require new (and probably difficult) proof techniques.
\item How hard are the fixed levels of $\QACC$? While lower bounds for 
$\QACC$
itself seem
impossible at present, it might be fruitful to study the limitations
of small depth $\QACC$  circuits (depth 2, for example).
\end{itemize}

\noindent {\bf Acknowledgments}: We thank Bill Gasarch for helpful
comments and suggestions.


\begin{thebibliography}{1}

\bibitem{ADH97} L.~Adleman, J.~DeMarrais, and M.~Huang.  Quantum
computability.  {\em SIAM J. Computing} {\bf 26} (1997) 1524--1540.

\bibitem{ajtai}
M. Ajtai.
``$\Sigma_1^1$ formulae on finite structures.''
{\em Ann. Pure Appl. Logic} {\bf 24} (1983) 1--48.

\bibitem{barenco95}
A.~Barenco, C.~Bennett, R.~Cleve, D.P.~DiVincenzo, N.~Margolus, P.~Shor,
T.~Sleator, J.A.~Smolin, and H. Weifurter,
``Elementary gates for quantum computation.''
{\em Phys. Rev. A} {\bf 52}, pages 3457--3467, 1995.

\bibitem{barrington} D. A. Barrington,
``Bounded-width polynomial-size branching programs recognize
exactly those languages in $\NC^1$.''
{\em J. Comput. System Sci.} {\bf 38} (1989) 150--164.

\bibitem{beaudry} M. Beaudry, P. McKenzie, P. P\'eladeau, and
D. Th\'erien, ``Circuits with monoidal gates.''  {\em Proc. Symposium
on Theoretical Aspects of Computer Science} (1993) 555--565.

\bibitem{bennett} C.H. Bennett,
``Time/space trade-offs for reversible computation.''
{\em SIAM J. Computing} {\bf 4(18)} (1989) 766--776.

\bibitem{chungyao77}
K.~C.~Chung and T.~H.~Yao,
``On Lattices Admitting Unique Lagrange Interpolations.''
{\em SIAM Journal of Numerical Analysis} {\bf 14} (1977) 735--743.

\bibitem{cirac} J.I. Cirac and P. Zoller,
``Quantum computers with cold trapped ions.''
{\em Phys. Rev. Lett.} {\bf 74} (1995) 4091--4094.

\bibitem{ds} D.P. DiVincenzo and P.W. Shor,
``Fault-tolerant error correction with efficient quantum codes.''
quant-ph/9605031.

\bibitem{clote93} P.~Clote, ``On polynomial Size Frege Proofs of
Certain Combinatorial Principles.''  In P.~Clote and J.~Krajicek,
editors, {\em Arithmetic, Proof Theory, and Computational Complexity},
Oxford (1993) 164--184.

\bibitem{fj99} L.~Fortnow and J. Rogers, ``Complexity
Limitations on Quantum Computation.'' {\em Proc. 13th
Conference on Computational Complexity} (1998) 202--209.

\bibitem{fghr98} S.~Fenner, F.~Green, S.~Homer, and R.~Pruim,
``Quantum NP is hard for PH.''  {\em Royal Society of London A}
(1999) 455, pp 3953 - 3966.

\bibitem{furst} M. Furst, J.B. Saxe, and M. Sipser,
``Parity, circuits, and the polynomial-time hierarchy.''
{\em Math. Syst. Theory} {\bf 17} (1984) 13--27.

\bibitem{gershenfeld} N. Gershenfeld and I. Chuang,
``Bulk spin resonance quantum computation.''
{\em Science} {\bf 275} (1997) 350.

\bibitem{ghp00} F.~Green, S.~Homer, and C.~Pollett,
``On the Complexity of
Quantum ACC," in {\em Fifteenth Annual Conference on Computational
Complexity Theory,} IEEE Computer Society Press, (2000) 250-262.

\bibitem{moore99} C.~Moore, ``Quantum Circuits: Fanout,
Parity, and Counting.'' quant-ph/9903046.

\bibitem{circuits} C. Moore, D. Th\'erien, F. Lemieux, J. Berman, and
A. Drisko\ , ``Circuits and Expressions with Non-Associative Gates.''
{\em Journal of Computer and System Sciences} {\bf 60} (2000) 368--394.


\bibitem{codes} C. Moore and M. Nilsson, ``Parallel quantum
computation and quantum codes.''  quant-ph/9808027, and to appear in
{\em SIAM J. Computing}.

\bibitem{early} C. Moore and M. Nilsson,
``Some notes on parallel quantum computation.''
quant-ph/9804034.

\bibitem{nic72} R.~A.~Nicolaides, ``On a class of finite elements
generated by Lagrange interpolation.''  {\em SIAM Journal of Numerical
Analysis} {\bf 9} (1972) 177--199.

\bibitem{papa} C.H. Papadimitriou,
{\em Computational Complexity.}
Addison-Wesley, 1994.

\bibitem{razborov} A.A. Razborov,
``Lower bounds for the size of circuits of bounded depth with basis
$\{\&, \oplus\}$.''
{\em Math. Notes Acad. Sci. USSR} {\bf 41(4)} (1987) 333--338.

\bibitem{reck} M. Reck, A. Zeilinger, H.J. Bernstein, and P. Bertani,
``Experimental realization of any discrete unitary operator.''
{\em Phys. Rev. Lett.} {\bf 73} (1994) 58-61.

\bibitem{saks} M. Saks, ``Randomization and derandomization in
space-bounded computation.''  {\em Proc. 11th IEE Conference on
Computational Complexity} (1996) 128--149.

\bibitem{shor1} P.W. Shor, ``Algorithms for quantum computation:
discrete logarithms and factoring.''  {\em Proc. 35th IEEE Symposium
on Foundations of Computer Science} (1994) 124--134.

\bibitem{shor2} P.W. Shor,
``Fault-tolerant quantum computation.''
quant-ph/9605011.

\bibitem{shor97} P.~W. Shor, ``Polynomial-time algorithms for prime
number factorization and discrete logarithms on a quantum computer.''
{\em SIAM J. Computing} {\bf 26} (1997) 1484--1509.

\bibitem{siu91} K.-Y.~Siu and V~Rowchowdhury, ``On optimal depth
threshold circuits for multiplication and related problems.''  {\em
SIAM J. Discrete Math.} {\bf 7} (1994) 284--292.

\bibitem{siu94} K.-Y.~Siu and J~Bruck, ``On the power of threshold
circuits with small weights.''  {\em SIAM J. Discrete Math.} {\bf 4}
(1991) 423--435.

\bibitem{smo87} R.~Smolensky, ``Algebraic methods in the theory of
lower bounds for Boolean circuit complexity.''  {\em Proc. 19th Annual
ACM Symposium on Theory of Computing} (1987) 77--82.

\bibitem{yy98}
T. Yamakami and A.C. Yao, ``$NQP_{\bf C}= co$-$C_=P$.''
To appear in {\it Information Processing Letters}.

\bibitem{yao90} A.~C.-C. Yao, ``On ACC and threshold circuits.'' {\em
Proc. 31st IEEE Symposium on Foundations of Computer Science}
(1990) 619--627.

\bibitem{yao93}
A.~C.-C. Yao, ``Quantum circuit complexity.''
{\em Proc. 34th IEEE Symposium on Foundations of
Computer Science} (1993) 352--361.

\end{thebibliography}
\end{document}